\documentclass[twocolumn]{aastex62}
\usepackage{natbib}
\graphicspath{{./}{figures/}}
\usepackage{color,subfigure,lineno} 

\def\OIII{[O\,{\sc iii}]\,5007}
\def\ASTRID{{\tt ASTRID}}

\shorttitle{Galactic-Scale Quasar Pairs}
\shortauthors{Shen et~al.}

\begin{document}

\title{Statistics of Galactic-Scale Quasar Pairs at Cosmic Noon}


\author[0000-0003-1659-7035]{Yue Shen}
\affiliation{Department of Astronomy, University of Illinois at Urbana-Champaign, Urbana, IL 61801, USA}
\affiliation{National Center for Supercomputing Applications, University of Illinois at Urbana-Champaign, Urbana, IL 61801, USA}

\author[0000-0003-4250-4437]{Hsiang-Chih Hwang}
\affiliation{School of Natural Sciences, Institute for Advanced Study, Princeton, 1 Einstein Drive, NJ 08540, USA}

\author{Masamune Oguri}
\affiliation{Center for Frontier Science, Chiba University, 1-33 Yayoi-cho, Inage-ku, Chiba 263-8522, Japan}
\affiliation{Department of Physics, Graduate School of Science, Chiba University, 1-33 Yayoi-Cho, Inage-Ku, Chiba 263-8522, Japan}

\author{Nianyi Chen}
\affiliation{McWilliams Center for Cosmology, Department of Physics, Carnegie Mellon University, Pittsburgh, PA 15213, USA}

\author{Tiziana Di Matteo}
\affiliation{McWilliams Center for Cosmology, Department of Physics, Carnegie Mellon University, Pittsburgh, PA 15213, USA}
\affiliation{NSF AI Planning Institute for Physics of the Future, 
Carnegie Mellon  University, Pittsburgh, PA 15213, USA}

\author{Yueying Ni}
\affiliation{McWilliams Center for Cosmology, Department of Physics, Carnegie Mellon University, Pittsburgh, PA 15213, USA}
\affiliation{NSF AI Planning Institute for Physics of the Future, 
Carnegie Mellon  University, Pittsburgh, PA 15213, USA}

\author{Simeon Bird}
\affiliation{Department of Physics \& Astronomy, University of California, Riverside, 900 University Ave., Riverside, CA 92521, USA}

\author[0000-0001-6100-6869]{Nadia Zakamska}
\affiliation{Department of Physics and Astronomy, Johns Hopkins University}
\affiliation{School of Natural Sciences, Institute for Advanced Study, Princeton, 1 Einstein Drive, NJ 08540, USA}

\author[0000-0003-0049-5210]{Xin Liu}
\affiliation{Department of Astronomy, University of Illinois at Urbana-Champaign, Urbana, IL 61801, USA}
\affiliation{National Center for Supercomputing Applications, University of Illinois at Urbana-Champaign, Urbana, IL 61801, USA}

\author{Yu-Ching Chen}
\affiliation{Department of Astronomy, University of Illinois at Urbana-Champaign, Urbana, IL 61801, USA}

\author[0000-0001-5253-1338]{Kaitlin M. Kratter}
\affiliation{University of Arizona, 933 N Cherry Ave, Tucson, AZ 85721, USA}


\begin{abstract}
The statistics of galactic-scale quasar pairs can elucidate our understanding of the dynamical evolution of supermassive black hole (SMBH) pairs, the duty cycles of quasar activity in mergers, or even the nature of dark matter, but have been challenging to measure at cosmic noon, the prime epoch of massive galaxy and SMBH formation. Here we measure a double quasar fraction of $\sim 6.2\pm0.5\times 10^{-4}$ integrated over $\sim 0\farcs3-3\arcsec$ separations (projected physical separations of $\sim 3-30\,{\rm kpc}$ at $z\sim 2$) in luminous ($L_{\rm bol}>10^{45.8}\,{\rm erg\,s^{-1}}$) unobscured quasars at $1.5<z<3.5$, using Gaia EDR3-resolved pairs around SDSS DR16 quasars. The measurement was based on a sample of 60 Gaia-resolved double quasars (out of 487 Gaia pairs dominated by quasar+star superpositions) at these separations, corrected for pair completeness in Gaia, which we quantify as functions of pair separation, magnitude of the primary, and magnitude contrast. The double quasar fraction increases towards smaller separations by a factor of $\sim5$ over these scales. The division between physical quasar pairs and lensed quasars in our sample is currently unknown, requiring dedicated follow-up observations (in particular, deep, sub-arcsec-resolution IR imaging for the closest pairs). Intriguingly, at this point the observed pair statistics are in rough agreement with theoretical predictions both for the lensed quasar population in mock catalogs and for dual quasars in cosmological hydrodynamic simulations. Upcoming wide-field imaging/spectroscopic space missions such as Euclid, CSST and Roman, combined with targeted follow-up observations, will conclusively measure the abundances and host galaxy properties of galactic-scale quasar pairs, offset AGNs, and sub-arcsec lensed quasars across cosmic time. 
\end{abstract}



\section{Introduction}\label{sec:intro}

The formation of binary SMBHs ($M_{\rm BH}\gtrsim 10^6\,M_\odot$) is the inevitable consequence of galaxy mergers and the prevalence of SMBHs in galactic nuclei \citep[e.g.,][]{Begelman_etal_1980}. After the merger of two galaxies, the two SMBHs will in-spiral in the merged galaxy due to dynamical friction from tens of kpc to $\sim 10\,$parsec \citep[e.g.,][]{milosavljevic01,Yu_2002,DEGN}. The two SMBHs become a gravitationally bound binary at $\lesssim 10$\,parsec separations, and interactions with stars continue to shrink the binary orbit. The evolution of the binary SMBH below $\sim 1$\,parsec depends on the properties of stellar orbits in the galactic potential and the effects of gas \citep[e.g.,][]{DEGN,DeRosa_etal_2019}. But if the binary orbit can shrink to scales $\ll 1$\,parsec, gravitational wave (GW) radiation will take over in shrinking the binary orbit and eventually lead to the coalescence of the two SMBHs. The GW signals during the final in-spiral and coalescence of the binary SMBH are highly anticipated from ongoing pulsar timing arrays \citep[e.g.,][]{pta20} and future GW facilities such as the Laser Interferometer Space Antenna \citep[e.g.,][]{lisa22}. 

SMBH pairs at galactic-scale separations (tens of kpc to tens of parsec) represent the best-understood stage in theoretical studies of binary SMBH formation. The abundance of these wide separation pairs sets the initial conditions of binary SMBHs expected from galaxy mergers. Their pair separation statistics constrain the evolutionary timescales of galactic-scale SMBH pairs, which can be compared with analytical calculations or numerical simulations \citep[e.g.,][and references therein]{DEGN,DeRosa_etal_2019}. Dynamical friction dominates the orbital evolution of these pairs before they become bound binaries. Nevertheless, there are still lingering theoretical uncertainties in this regime, and the timescale spent at these galactic-scale separations depends on the galaxy potential, mass ratio of the merging galaxies, properties of the stellar cores surrounding each SMBH, as well as the effects of gas (both dynamical and accretion onto SMBHs) and dark matter halo properties \citep[e.g.,][]{milosavljevic01,Yu_2002,Callegari_etal_2009,Callegari_etal_2011,Khan2013,McWilliams_etal_2014,Kelley_etal_2017,Tremmel2018,Tamfal_etal_2018,ChenN_etal_2022}. 

Observationally, galactic-scale SMBH pairs can be identified as dual Active Galactic Nuclei (AGNs) or luminous dual quasars (conventionally defined by $L_{\rm bol}\gtrsim 10^{45}\,{\rm erg\,s^{-1}}$), if both SMBHs are active. Inactive SMBH pairs on galactic-scales are difficult to identify at cosmological distances. Pairs with only one active SMBH may appear as an offset AGN \citep[e.g.,][]{Barrows_etal_2016}. But the detection of offset AGNs becomes challenging at $z>1$, requiring deep imaging/spectroscopy and robust measurements of the host galaxy centroid, as well as careful treatments of selection effects \citep[e.g.,][]{Stemo_etal_2021}.

These dual AGNs/quasars signpost galactic-scale SMBH pairs, and can be used to constrain the underlying SMBH pair population, if the AGN duty cycle can be reliably inferred from hydrodynamic simulations. With sufficient statistics to explore the diversity of the SMBH pair population, such as host galaxy properties and redshift evolution, observations of these pairs will enable critical comparisons with theoretical models.  

The pairing and dynamical evolution of SMBHs at $z\sim 2$ is of particular importance. The specific galaxy merger rate is much higher at cosmic noon than at lower redshifts \citep[e.g.,][]{Duncan_etal_2019}, where both luminous quasars and global star formation reached their peak activity around $z\sim 2$ \citep[e.g.,][]{Madau_Dickinson_2014,Richards_etal_2006a}. This is the prime epoch of the growth of massive SMBHs and galaxies, and the onset of formation of the most massive (e.g., $>10^8\,M_\odot$) SMBH binaries, whose eventual coalescence will dominate the GW signal in the pulsar timing array band. The statistics of galactic-scale SMBH pairs at cosmic noon, as traced by dual quasars, provide critical constraints on the dynamical friction timescales, as well as the impact of galaxy mergers on the fueling of SMBHs. 

The pair statistics down to $\sim 1$\,kpc may even constrain the nature of dark matter. For example, in the fuzzy dark matter model \citep{Hu2000} and neglecting baryonic effects, SMBH pairs would never get much closer than $\sim$1 kpc because fuzzy dark matter fluctuations inhibit the orbital decay and inspiral at $\sim$~kpc scales \citep{Hui2017}, resulting in a ``pile up'' of SMBH pairs at $\sim 1$~kpc. A spike in the dual quasar fraction towards $\sim 1\,$kpc, above the level that can be explained by quasar duty cycle enhancement in mergers, may be the smoking gun signature of fuzzy dark matter.  

Unfortunately, given the stringent spatial resolution requirement (e.g., sub-arcsec for $\sim$kpc scales) and the apparent rareness of such pairs, the observational inventory of $z>1.5$ dual quasars at $\lesssim$ tens of kpc separations remains scarce. There are only a handful of serendipitously discovered $\sim$kpc-scale dual/offset AGNs known at lower redshifts \citep[e.g.,][]{Komossa_etal_2003,Comerford_etal_2009b,Liu_etal_2010b,Civano_etal_2010,Goulding_etal_2019}. Dedicated wide-area searches of binary quasars\footnote{For historical reasons, these wide-separation pairs are referred to as ``binary quasars'' \citep[e.g.,][]{Djorgovski_1991,Kochanek_etal_1999} as the two SMBH+galaxy systems are bound to each other. } at $z>1.5$ have compiled tens of quasar pairs at projected separations of $10\,{\rm kpc}\lesssim r_p\lesssim 50\,{\rm kpc}$  \citep[e.g.,][]{Hennawi_etal_2006,Hennawi_etal_2010,Myers_etal_2008,Kayo_Oguri_2012,More_etal_2016,Eftekharzadeh_etal_2017}, starting to probe the galactic-scale environment of quasar pairs. But the $r_p<10\,$kpc regime of high-redshift quasar pairs remains largely unexplored \citep[fig.~1 in][]{Chen_etal_2022}, due to the lack of efficient quasar identification for sub-arcsec pairs that are typically unresolved in ground-based data. Assuming no merger-enhanced AGN duty cycles and applying dynamical friction prediction of galactic-inspiral timescales \citep[i.e., the dynamical friction timescale $t_{\rm df}$ is roughly proportional to $r$, the 3D pair separation, e.g.,][]{Yu_2002,Chen_etal_2020b}, we expect a $\sim$kpc-scale dual quasar fraction of $f_{QQ}\sim 5\times 10^{-5}$ among all quasars, extrapolated from the observed quasar pair statistics on tens of kpc scales \citep[e.g., $f_{QQ}\sim 5\times10^{-4}$,][]{Kayo_Oguri_2012}. To test these expectations, we need to search a large parent quasar sample in order to build up the statistics of rare dual quasars. 

In this work we measure the galactic-scale (i.e., $r_p\lesssim 30\,{\rm kpc}$) quasar pair fraction at $z\sim 2$ using a different approach than earlier studies \cite[e.g.,][]{Hennawi_etal_2006,Hennawi_etal_2010,Myers_etal_2008,Kayo_Oguri_2012,More_etal_2016,Eftekharzadeh_etal_2017,Silverman_etal_2020}, focusing on the $r_p<10$\,kpc regime that has been poorly explored before. Our approach builds on the all-sky Gaia survey Early Data Release 3 \citep{Fabricius_etal_2021}, which provides precise coordinates, magnitudes, and astrometric measurements for all-sky sources to as faint as $G\sim 21$. In particular, Gaia's nominal $\sim 0\farcs2$ resolution enables the identification of close-separation companions around distant quasars, with quantifiable completeness in resolved pairs as a function of angular separation (\S\ref{sec:data}). Importantly, Gaia proper motion measurements enable efficient separation of stars and quasars, a unique advantage that previous quasar pair searches based on photometric color selection did not have. There is no need to update our analysis using the recent Gaia DR3 release \citep{gaiadr3} since the photometric and astrometric content is essentially unchanged from EDR3 to DR3. 


This paper is organized as follows. In \S\ref{sec:data} we describe the sample and data used in our systematic search of high-redshift small-scale quasar pairs, with an emphasis on quantifying the completeness of Gaia EDR3 resolved pairs. We present our results in \S\ref{sec:result}, where we compare the observed pair statistics with theoretical predictions of lensed quasars and quasar pairs. We discuss the implications of our findings in \S\ref{sec:disc} and summarize in \S\ref{sec:con}. In this work, we focus on luminous unobscured broad-line quasars exclusively, given the survey depth of Gaia. Occasionally we use the term ``dual quasars'' to refer to physical quasar pairs on galactic scales, following the convention for dual AGNs at $z<1$ \citep[e.g.,][]{Comerford_etal_2009b} that have much lower luminosities than our quasars. By default quasar pairs refer to physically associated pairs within the merging galaxies, rather than unrelated, projected quasar pairs at different redshifts. For practical purposes, we use the term ``double quasars'' to collectively refer to quasar pairs and lensed quasars. We adopt a flat $\Lambda$CDM cosmology with $\Omega_\Lambda=0.7$, $\Omega_{M}=0.3$ and $H_0=70\,{\rm km\,s^{-1}\,Mpc^{-1}}$. Pair physical separations are measured in proper units. 


\section{Data}\label{sec:data}

We start from the latest compilation of spectroscopically confirmed quasars in SDSS-DR16 \citep[DR16Q,][]{Lyke_etal_2020}, and restrict our search to $z>1.5$ quasars. This redshift cut is crucial to this study, and ensures negligible emission from the host galaxy within the Gaia bandpass, which would complicate the source detection and astrometry measurements \citep{Hwang_etal_2020}. We then search for Gaia EDR3 sources in a 3\arcsec\ radius circular region around each SDSS quasar. We further require the matched Gaia sources to have $G<20.25$, which balances the needs for pair statistics and high completeness in Gaia detection and astrometric measurements. For example, \citet{Fabricius_etal_2021} demonstrated nearly 100\% completeness of photometric detection at $G=20$ in low stellar density fields with Gaia EDR3, applicable to SDSS quasars. We have tested Gaia's photometric detection completeness for single sources using the DR16Q quasar catalog, and find that the completeness is $\sim 98.12\pm 0.41\%$ even in the faintest bin $G=[20,20.25]$.

Our $G<20.25$ flux limit roughly corresponds to bolometric luminosity $L_{\rm bol}>10^{45.8}\,{\rm erg\,s^{-1}}$ at $z>1.5$ \citep{Shen_etal_2011}, or SDSS $i< 20.13$ (we adopt a magnitude conversion of $G=i+0.12$ assuming a fixed quasar power-law continuum $f_\nu\propto \nu^{-0.5}$). The parent sample satisfying these redshift and magnitude cuts and having single Gaia matches includes 134,796 DR16Q quasars. 

We focus on Gaia resolved double sources at the SDSS quasar position. Multiple systems with more than two Gaia sources brighter than $G=20.25$ within 3\arcsec\ are only $\sim 2\%$ of double systems, hence negligible. A more important issue is that the completeness of these multiples is much lower and much harder to quantify; thus we ignore this higher-order multiple population. Some quasars with only one matched Gaia source may still be a sub-arcsec quasar pair, which can be recovered with other approaches using additional Gaia parameters \citep[e.g.,][]{Hwang_etal_2020,Shen_etal_2019d,Chen_etal_2022,Mannucci_etal_2022,Makarov_Secrest_2022}, but are not covered here; instead, their contribution to the pair statistics is estimated through the completeness analysis (\S\ref{sec:pair_comp}). 

\subsection{The Pair Sample}\label{sec:sample}



Our initial Gaia-resolved pair sample includes 497 SDSS-DR16Q quasars. However, in 10 pairs both components are bona fide quasars listed in DR16Q and thus are counted twice. Removing these 10 duplicated pairs, our final Gaia-resolved pair sample includes 487 unique pairs. For each pair, the closer Gaia match is designated as the corresponding SDSS DR16Q quasar. This is generally the case. However, in very rare cases of pairs separated by $\lesssim 1\arcsec$, the SDSS optical centroid may be dominated by the companion. Nevertheless, this detail does not affect any of our statistical analyses below. We classify the companion as ``star-like'' in 416 pairs where its proper motion is detected by Gaia at $>3\sigma$ significance; for comparison, only $\sim 2\%$ of Gaia singly-matched quasars have $>3\sigma$ proper motion detection, meaning our proper motion cut will only inadvertently exclude a negligible fraction of bona fide double quasars. The remaining 71 resolved pairs are our initial sample of double quasars. Pair separations are computed using Gaia EDR3 coordinates, which can slightly exceed the 3\arcsec\ cross-matching radius between SDSS and Gaia. The full pair catalog of 487 pairs is presented in Table \ref{tab:sample}. 



\begin{figure*}
  \includegraphics[width=0.98\textwidth]{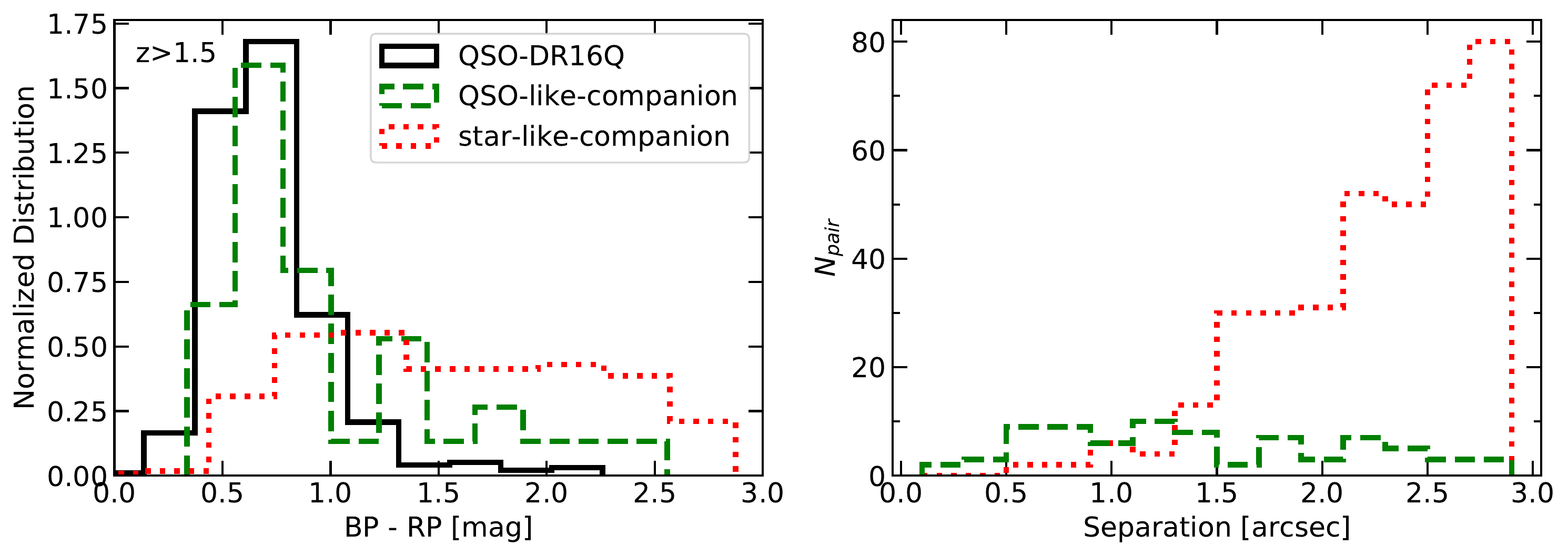}
  \caption{Distributions of the parent pair sample in Gaia $BP-RP$ color (left) and separation (right). The color distributions of DR16Q quasars (``QSO-DR16Q'') and companions classified as ``QSO-like'' based on proper motion detection are markedly different from that of companions classified as ``star-like''. The pair separation distributions in the right panel are also different for pairs with ``QSO-like'' and ``star-like'' companions. In particular, the pair separation distribution for ``star-like'' companions drops rapidly towards smaller separation, as expected from the reduction of the geometric cross section of foreground superpositions. In contrast, the pair separation distribution for ``QSO-like'' companions is more or less flat across these separations. These distributions are based on the raw pair statistics, without corrections for pair completeness towards the sub-arcsec regime (see \S\ref{sec:pair_comp}) or removal of the 10 star-quasar superpositions in $<1\farcs5$ pairs based on the PCA analysis (\S\ref{sec:data}).
  \label{fig:rawdist}}
  \end{figure*}

Fig.~\ref{fig:rawdist} (left) shows the distributions of Gaia $BP-RP$ color for the DR16Q quasar in the pair, ``star-like'' and ``quasar-like'' companions for the full Gaia pair sample. Because Gaia photometry is measured within a $3\farcs5\times 2\farcs1$ window \citep{Riello2021}, source deblending may be significantly impacted for the closest pairs. Thus we have excluded pairs with separations $<1\arcsec$ in this color distribution plot to avoid crosstalks in their photometric color measurements. Their color distributions suggest that ``star-like'' companions indeed have different colors than the primary quasars or the ``quasar-like'' companions based on Gaia proper motion detection.

\begin{table}
\caption{Pair Sample Data}\label{tab:sample}
\resizebox{\columnwidth}{!}{%
\begin{tabular}{llll}
\hline\hline
Column & Format & Units & Description \\
(1) & (2) & (3) & (4) \\
\hline
SDSS\_NAME  &     STRING  & &   J2000 hhmmss.ss$\pm$ddmmss.s   \\
   Z           &    DOUBLE         & &  Default redshift from DR16Q \\
   PLATE     &      LONG       & &      Plate number (SDSS spec) \\
   FIBERID     &    LONG   & &            FiberID (SDSS spec) \\
   MJD          &   LONG        & &     MJD (SDSS spec) \\
   GAIA\_RA1      &   DOUBLE     & deg &     Gaia RA \\
   GAIA\_DEC1     &  DOUBLE      & deg &      Gaia DEC \\
   GAIA\_RA2      &  DOUBLE    &deg &      Gaia RA\\
   GAIA\_DEC2    &   DOUBLE     &deg &      Gaia DEC \\
   G1    &          DOUBLE       & mag &    Gaia G mag\\
   G2    &          DOUBLE       & mag &     Gaia G mag \\
   BP\_RP1    &      DOUBLE    & mag &      Gaia BP-RP color \\
   BP\_RP2     &     DOUBLE   & mag &              Gaia BP-RP color \\
   PM\_SIG1   &      DOUBLE    &  &      PM significance\\
   PM\_SIG2    &     DOUBLE    &  &       PM significance\\
   PAIR\_SEP     &   DOUBLE     & arcsec &      Pair separation\\
   TYPE    &        STRING   &  &  Pair classification \\
   KNOWN &  STRING & & Literature classification \\
   F\_COMP & DOUBLE & & pair completeness (\S\ref{sec:pair_comp})\\
\hline
\hline\\
\end{tabular}
}
{\raggedright Notes. For each pair, index 1 refers to the DR16Q quasar and 2 refers to the companion, regardless of their relative brightness (i.e., the quasar can be fainter than the companion, especially at large pair separations). Gaia measurements are from EDR3 (null values are ``NaN''). The column ``TYPE'' indicates pair classification: ``QQ'' refers to double quasar; 
``QS\_PM'' refers to quasar+star pair based on proper motion; ``QS\_PCA'' refers to quasar+star pair based on spectral PCA analysis; one quasar (J0033+2015) is a known quasar+star pair \citep{More_etal_2016} and we set its TYPE=``QS\_KNOWN''. The associated FITS file is available in the online version of this paper. }
\end{table}

Fig.~\ref{fig:rawdist} (right) shows the distributions of pair separation for double systems with ``star-like'' and ``quasar-like'' companions. The separation distribution for ``star-like'' companions rapidly decline towards smaller separations, as anticipated from the reduction of geometric cross section and the constant sky density of a foreground (star) population, modulo pair-resolving incompleteness towards $\lesssim 1\arcsec$ separations. In contrast, the separation distribution for ``quasar-like'' companion remains more or less constant, suggesting that it is an intrinsic population associated with the primary quasar. Both the color and separation distributions in Fig.~\ref{fig:rawdist} indicate that the classification of star and quasar companions based on proper motion is reasonably good. Of course, it is possible that some detected proper motions are caused by systematics (especially for sub-arcsec pairs where the two sources overlap in photometric/astrometric measurements). Here we opt to exclude these potential double quasars mis-identified as quasar+star pairs due to bad proper motion measurements, in order to maintain a high-purity double quasar sample.  

\begin{figure*}
\centering
  \includegraphics[width=0.48\textwidth]{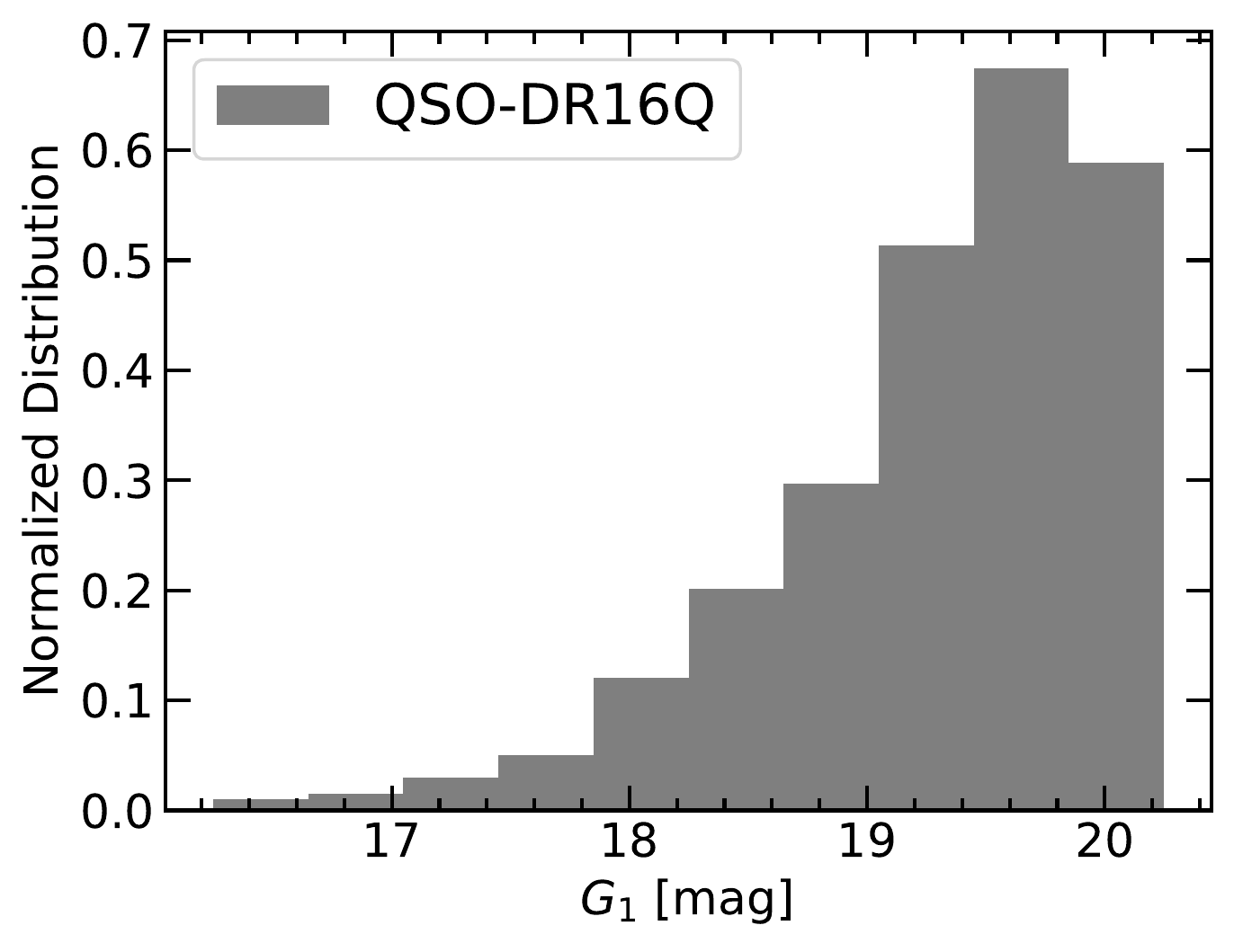}
  \includegraphics[width=0.48\textwidth]{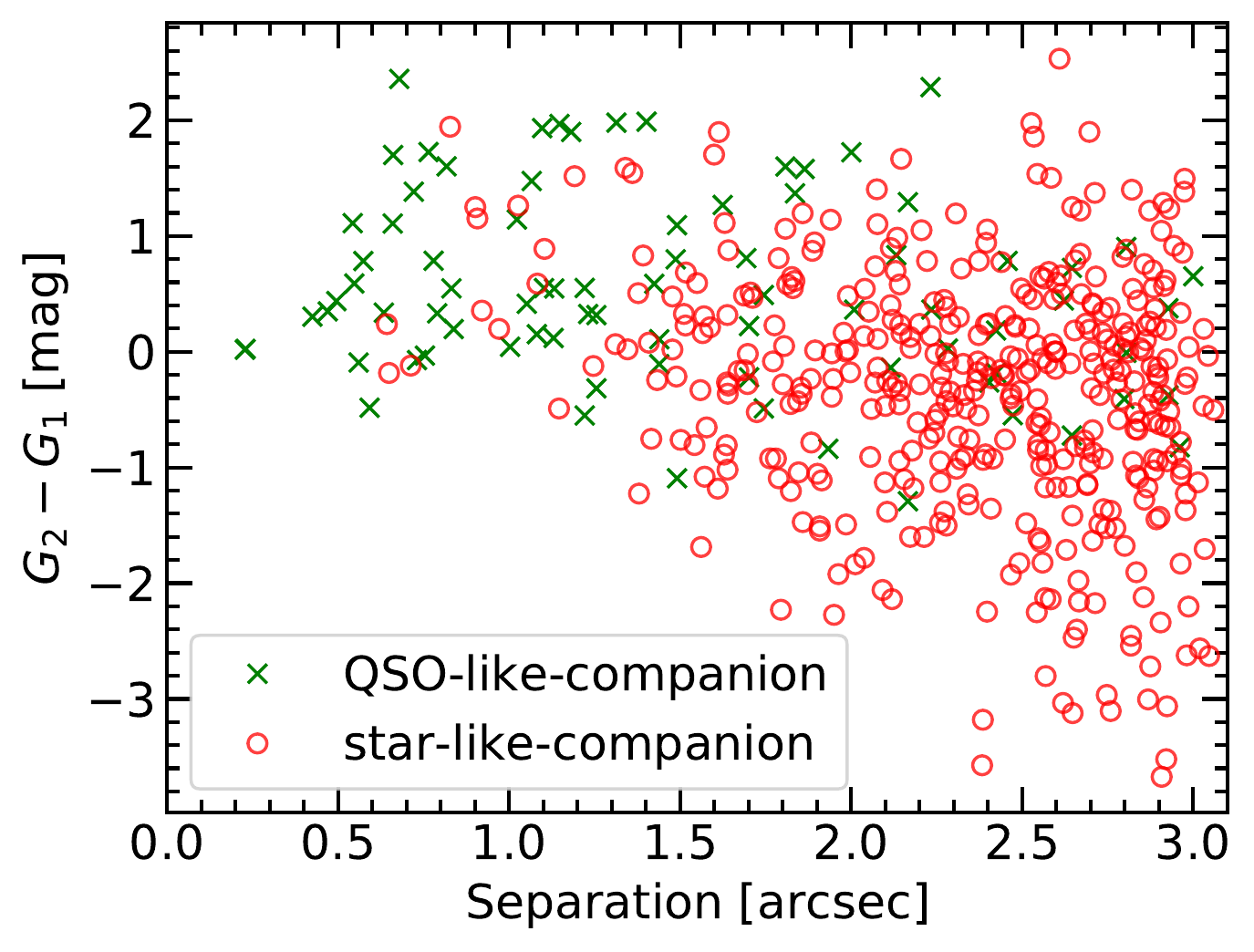}
  \caption{Statistical properties of the parent pair sample. {\em Left:} G-band magnitude distribution for the quasar component in DR16Q ($G_1$) in each pair. {\em Right:} magnitude contrast ($G_2-G_1$) between the companion and the DR16Q quasar as a function of pair separation. At large separations, the companion can be much brighter than the DR16Q quasar, especially in the case of star-like companions. However, the vast majority of pairs have flux contrast less than a factor of 10. As in Fig.~\ref{fig:rawdist}, the pairs here have not been corrected for completeness in the sub-arcsec regime, or cleaned based on the PCA analysis. 
    \label{fig:pair_prop}}
  \end{figure*}

\begin{figure*}
\centering
  \includegraphics[width=0.95\textwidth]{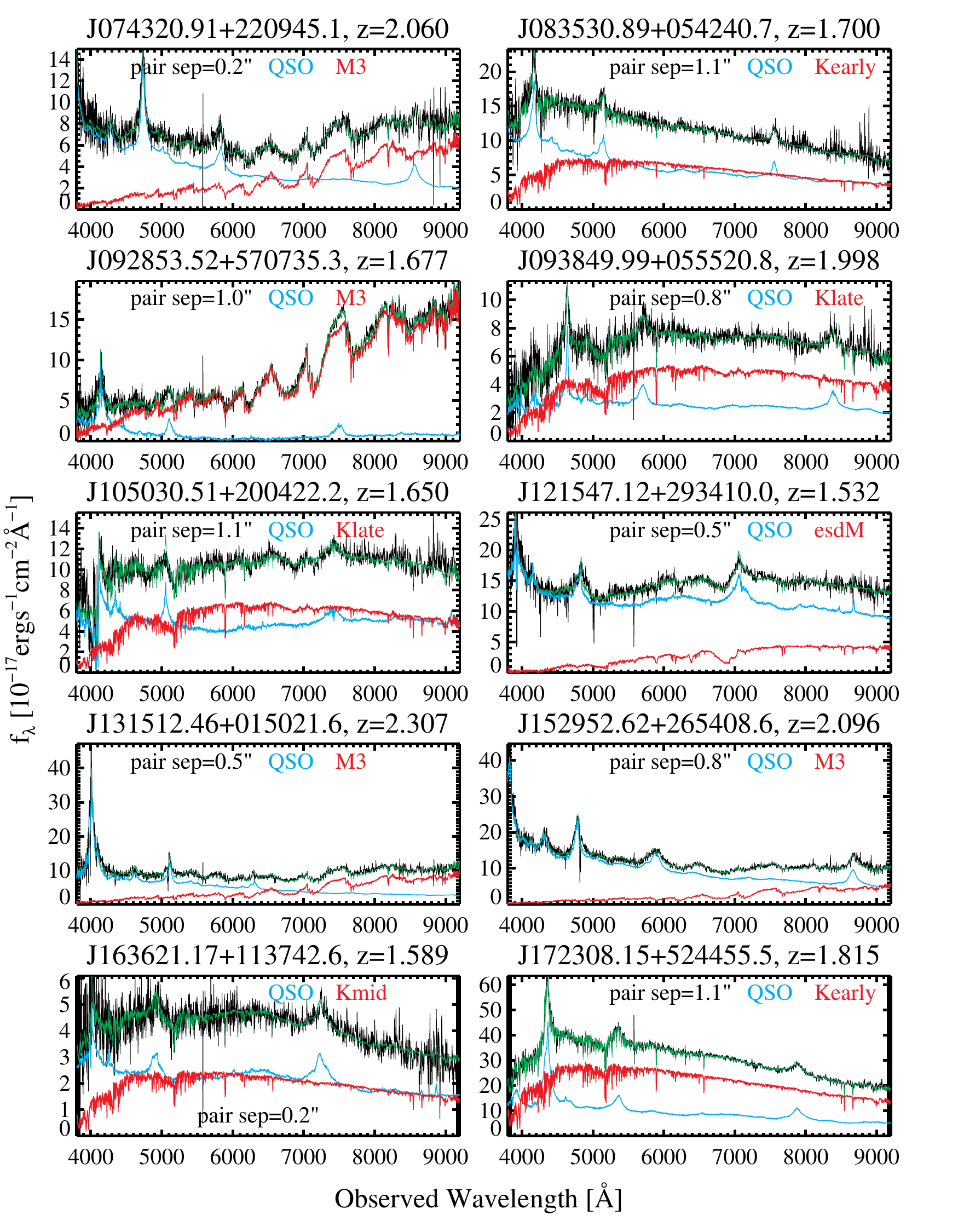}
  \caption{Spectral PCA results for the 10 pairs with separations $<1\farcs5$ originally included in our double quasar sample, but are best explained by star+quasar superpositions based on the SDSS spectra (\S\ref{sec:data}). In each panel, the black and green lines are the raw spectrum and the reconstructed spectrum, respectively. The cyan and red lines are the decomposed components from the quasar templates and the stellar templates, respectively, with the stellar type indicated in the upper right corner.  
    \label{fig:pca}}
  \end{figure*}

Likewise, we expect that there is still residual contamination of star superposition in these close pairs which we classified as ``quasar-like'' companions, especially at $\lesssim 1\arcsec$ separations where the measurement of Gaia proper motion is either unavailable or could be impacted by the close neighbor. We estimate this residual contamination rate using 43 pairs at $<1\farcs5$ separations from the initial sample of 71 double quasars. These pairs are close enough such that the SDSS fiber spectroscopy (with a fiber diameter of 2\arcsec\ or 3\arcsec) encloses most light from both components. We use a spectral Principal Component Analysis (PCA) technique to decompose the SDSS spectrum into potential quasar+star superpositions, using quasar and stellar PCA templates from the SDSS website. Fig.~\ref{fig:pca} shows that such superpositions can be reliably identified from the SDSS spectrum, provided that the companion is not substantially fainter (e.g., by a factor of $\sim 10$ in flux) than the primary quasar (96.6\% of Gaia resolved pairs in our sample have flux contrast ratio $<10$). However, automatic classifications with PCA-decomposed spectra are often unreliable due to degeneracies in the decomposition and noise in the data. Therefore we manually inspect all PCA decomposition results and flag obvious star superpositions. 

This spectral analysis indicates that there is $\sim 23\%$ (10/43 in the subset of $<1\farcs5$ pairs) contamination of star+quasar superposition in this subset of pairs. These apparent star superpositions have separations between 0\farcs2 and 1\farcs2, with no obvious dependence of the contamination rate on pair separation given the small number statistics. The PCA results for these 10 apparent quasar-star superpositions are shown in Fig.~\ref{fig:pca}. We remove these apparent quasar+star pairs from our double quasar sample. There is no way to remove additional stellar contamination in the $>1\farcs5$ pairs without additional follow-up observations. However, the proper motion measurements are much more reliable for pairs separated by $>1\arcsec$ to remove star superpositions in our initial cut. Thus we expect the residual contamination rate is substantially smaller than $\sim 20\%$ at $>1\farcs5$ separations.

The same spectral analysis reveals no obvious, physically unrelated, projected quasar pairs, in which case we would observe different emission line redshifts in the spectrum if the redshift difference is $>2000\,{\rm km\,s^{-1}}$ \citep[the common definition of projected quasar pairs, e.g.,][]{Hennawi_etal_2006,Hennawi_etal_2010}. This is consistent with our expectation from reduced cross section of chance superpositions for our $<3\arcsec$ separation pairs: \citet{Hennawi_etal_2010} estimated $\sim 30\%$ of the double quasars at $<60\arcsec$ separations are projected pairs, which would imply negligible projected pairs at $<3\arcsec$ separations. 

After removing foreground star superpositions, most of the remaining 61 pairs should either be genuine dual quasars, or gravitationally lensed quasar images. Extended host galaxy emission from old stellar populations at $z>1.5$ would be too faint to be detectable in the Gaia band, and compact UV-emitting star formation regions in the host galaxy is unlikely to be brighter than our flux limit (which implies quasar luminosities). However, the population of lensed quasars cannot be readily removed. Indeed, resolved Gaia pairs have been used to identify candidate gravitationally lensed quasars and confirmed in follow-up observations \citep[e.g.,][]{Lemon_etal_2017,Lemon_etal_2018,Lemon_etal_2022,Krone-Martins_etal_2018}. We cross-match the 61 pairs in our sample with the Gravitationally Lensed Quasar Database\footnote{https://research.ast.cam.ac.uk/lensedquasars/; latest version in 2019.} and the follow-up sample of Gaia DR2-selected candidate lenses and quasar pairs in \citet{Lemon_etal_2022}, as well as additional SDSS quasar lens and pair searches \citep[][]{Hennawi_etal_2006,Hennawi_etal_2010,Oguri_etal_2008,Inada_etal_2012,Myers_etal_2008,Kayo_Oguri_2012,More_etal_2016,Eftekharzadeh_etal_2017}. We find that there are 25 systems that are reported lenses (but only four of them have image separations $<1\arcsec$) in follow-up observations. There are 5 systems reported as a physical quasar pair. These publicly reported cases are indicated in the ``KNOWN'' column in Table~\ref{tab:sample}. It is possible that there are additional sources observed in the literature that are missed from the above resources. Mis-classifications of lenses and pairs among these reported cases are rare but possible (see discussions in \S\ref{sec:disc1}). Finally, three additional pairs among the 61 (J082341.08$+$241805.6, J084129.77$+$482548.4, and J212243.01$-$002653.8) have been observed in our pilot follow-up with HST (optical and IR) and/or VLA. J0823 and J0841 are confirmed double quasars, more likely dual than lensed quasars (Y.~Chen et~al., in prep.). J2122 was reported as a dual/lensed quasar based on resolved 2-band optical HST color \citep[][]{Chen_etal_2022}, pending further confirmation from additional follow-ups. 

During cross-matching our full SDSS+Gaia pair sample (487 pairs) with the above literature on quasar pairs and lenses (as well as our ongoing follow-up), we found one system (J135306.34+113804.7) classified by us as a quasar-star pair based on Gaia proper motion turns out to be a lensed quasar \citep{Inada_etal_2012}. On the other hand, only one system (J003337.58+201538.1) classified by us as a double quasar (separated by 1\farcs69) turns out to be a quasar-star pair based on spatially resolved optical spectroscopy \citep{More_etal_2016}. We remove J0033+2015 from further analysis, leaving a final cleaned double quasar sample of 60 objects. 


\begin{figure}
\centering
  \includegraphics[width=0.48\textwidth]{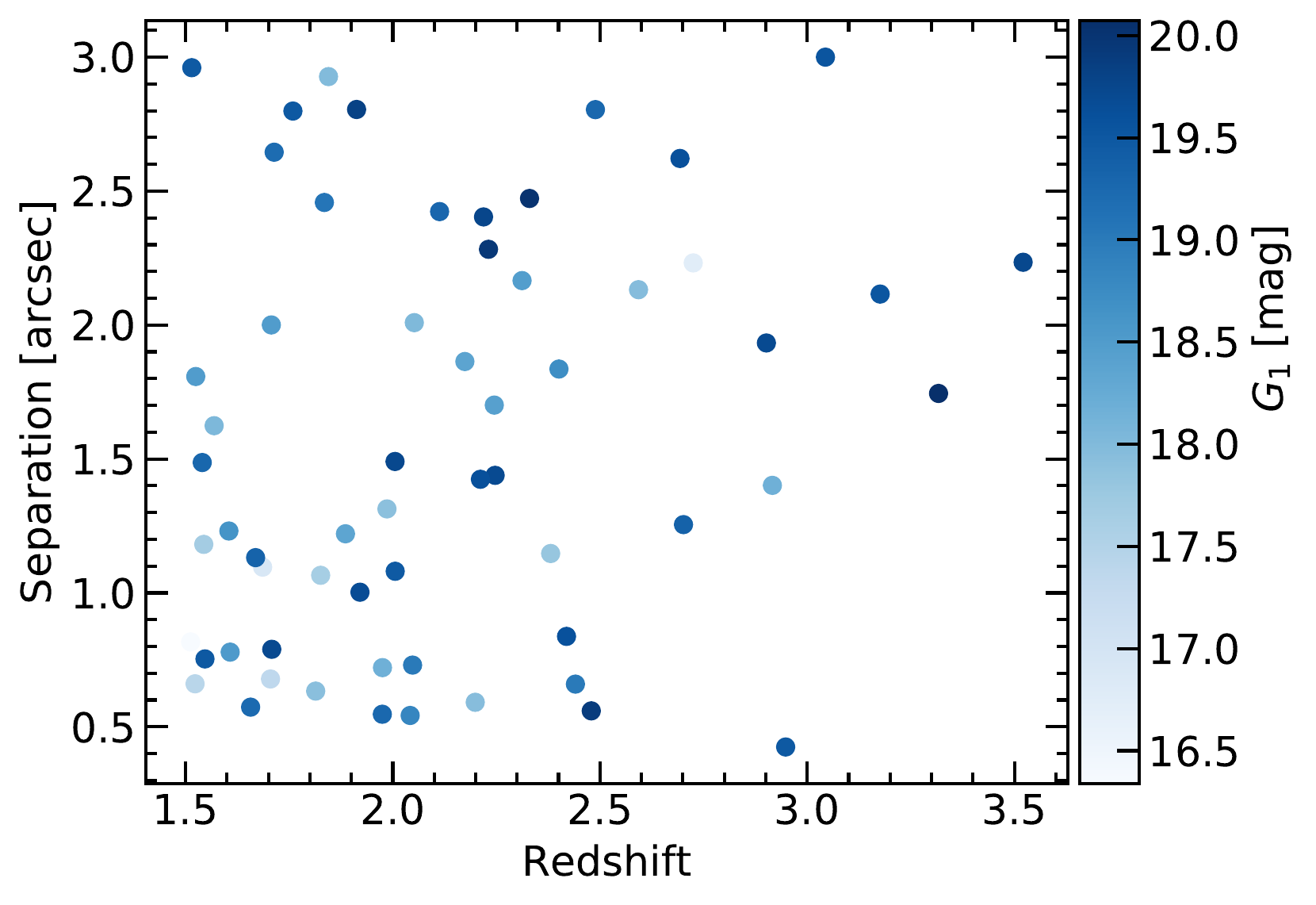}
  \caption{Distribution of our final sample of 60 double quasars in the redshift-separation space, color-coded by the $G$-band magnitude of the DR16Q member of each pair. 
    \label{fig:z_sep}}
  \end{figure}

Unfortunately, the completeness of follow-up observations of candidate quasar pairs or lensed quasars is difficult to quantify and varies across different surveys. Moreover, the constraints on the lensed quasar population in the sub-arcsec regime are essentially absent. For these reasons, we statistically evaluate the contribution of lensed quasars in our pair sample using mock catalogs, as described in \S\ref{sec:result}. Nevertheless, the fact that $\sim$half of our double quasar sample are already confirmed lensed quasars or quasar pairs indicates that our SDSS+Gaia selection is highly effective, and the resulting sample of 60 objects has a high purity of genuine double quasars. 

Fig.~\ref{fig:z_sep} displays the distribution of these 60 pairs in the redshift-separation space. These double quasars have pair separations between 0\farcs4 and $\sim 3\arcsec$, and form the basis of our subsequent analyses. At the sample median redshift of $z=2$, these pairs probe projected separations of $3\lesssim r_p\lesssim 30\,{\rm kpc}$, i.e., on galactic scales. Individual pairs may still have 3D separations exceeding 30\,kpc, but statistically this population still traces the radial distribution of quasar pairs. Projection effects are properly taken into account when comparing with theoretical predictions in \S\ref{sec:result} and \S\ref{sec:disc}. 

The fact that close photometric companions within 3\arcsec\ of SDSS quasars are dominated by the foreground (star) population signifies the necessity of additional metrics to remove foreground contamination in quasar pair searches. This high foreground contamination rate is verified in a random offset test. We shuffle the positions of SDSS quasars by 1 arcmin and search for Gaia sources within a 3\arcsec-radius circle. This random offset test maintains the foreground stellar density distribution applicable to the SDSS quasar sample. We find a chance star superposition around 0.68\% of the quasars when limiting to the same $G<20.25$ limit of Gaia sources. This is even higher than the $\sim 0.36\%$ rate for the observed sample above. The main reason for the lower superposition contamination rate in the observed quasar sample is due to SDSS selection. SDSS quasars were targeted by color selection with photometry (with $\sim 1\farcs4$ seeing) and spectroscopically confirmed with 2\arcsec\ or 3\arcsec-diameter fiber spectroscopy. The presence of a star brighter than the quasar itself will both impact the target selection and the spectroscopic classification. In this sense, the SDSS quasar sample is biased against close pairs with bright star companions. If we further require $G$ is no more than 2.5 magnitude brighter than the quasar in the offset test, we find 0.47\% quasars have chance foreground superpositions, roughly consistent with the observed rate. Additional effects, e.g., incompleteness in resolving pairs at $<1\arcsec$, would further reduce the observed foreground contamination rate.



\subsection{Pair Completeness}\label{sec:pair_comp}


The raw observed pair statistics as a function of separation (Fig.~\ref{fig:rawdist} right) suffer significantly from incompleteness in the sub-arcsec regime, as Gaia can only resolve the pair at $\sim 0\farcs2$ resolution in the along-scan direction (this is somewhat remedied by multiple scans along different directions). Moreover, the presence of a close neighbor decreases the probability of detecting both sources photometrically by Gaia.  The pair-resolving completeness as a function of separation has been estimated \citep{Fabricius_etal_2021} using the Washington Double Star Catalog (WDS) catalog \citep{Mason_etal_2001}, demonstrating significant improvements of EDR3 over DR2. Based on the WDS catalog, the pair completeness is $\sim 50\%$ (20\%) at 0\farcs5 (0\farcs3) separations. However, the WDS catalog has a different magnitude distribution (i.e., much brighter) than the parent SDSS quasar sample, and it is reasonable to expect that the completeness of Gaia-resolved pairs depends on both magnitude and magnitude contrast. Therefore, we carry out an independent measurement of the pair completeness in Gaia EDR3, as detailed below. 


We consider the detectability of close pairs as functions of the magnitude of the brighter primary source, the magnitude difference between the two sources, and their angular separation. There have been previous studies focusing on the Gaia completeness correction as functions of angular separation \citep{Fabricius_etal_2021} and magnitude differences \citep{El-Badry2018b}, but the exact completeness correction depends on the detailed selection criteria of the sample of interest \citep{El-Badry2018b}. 


We assemble a random pair sample where the pairs are dominated by random stellar pairs, and derive the completeness correction by comparing the observed number of pairs with the expected number of random pairs from a constant sky density of stars, $N ds \propto sds$, where $s$ is the projected angular separation. Following \cite{Hwang2022ecc}, we collect all pairs in the crowded field at $30^\circ<l<55^\circ$ and $5^\circ<b<7^\circ$. This region is chosen such that the Gaia source density is high at low Galactic latitudes, and the region is not strongly affected by dust extinction. We query all Gaia EDR3 sources within this sky region, without any other criteria. Then we collect all pairs with angular separations $<10$\arcsec. To reduce the binary star contribution, which is more prominent at $G<16$ (because brighter stars are closer and thus their binary companions are more likely to be spatially resolved), we further impose a cut on parallaxes $<0.5$\,mas, resulting in 16.7 million unique pairs.



We derive the completeness correction as a function of three parameters: magnitude of the brighter primary ($G_{pri}$), magnitude difference ($\Delta G = G_{sec}-G_{pri}$ where $G_{sec}$ is the $G$-band magnitude of the secondary), and angular separation. To this end, we bin these random pairs by $G_{pri}=15-21$ with steps of 1\,mag, $\Delta G=0-3$ with steps of 0.5\,mag, and angular separations $0\farcs 1-4\arcsec$ with steps of 0\farcs2. Then at each point of the 3D parameter grid, we compute the completeness factor $f_{\rm comp}=N_{obs}/N_{model}$, where $N_{obs}$ is the observed number of pairs in the grid and $N_{model}$ is the expected number of pairs from the model. 

The model is computed as follows. First, for every primary magnitude bin, we start with the first magnitude difference bin (i.e., $0<\Delta G<0.5$) and compute the expected number of pairs along the separation bin based on the observed number of pairs at 5-10\arcsec\ and the expected geometric distribution ($N ds \propto sds$). Next, under the assumption that the sample is dominated by random pairs and therefore the magnitude difference distribution is independent of pair separation, we use the magnitude difference distribution from pairs at 5-10\arcsec\ as the ground truth, and apply this distribution to smaller separations to obtain the expected pair counts as a function of $\Delta G$ at different separations.

Our completeness correction uses the expected geometric distribution $Nds\propto sds$, which is applicable when the sample is indeed dominated by random pairs in each 3D bin. While the overall sample is dominated by random star pairs in this crowded region \citep{Hwang2022ecc}, if we naively bin the sample into the 3D grid without the parallax cut, there are some noticeable binary contributions that cause the completeness correction $>1$ at $(G_{pri}<16) \wedge (\Delta G<0.5) \wedge$ (angular separations $<1$\arcsec), where $\wedge$ is the logical AND operator. This binary contribution in the observed data is due to the fact that brighter stars are closer, and thus their binary companions are more likely to be resolved by Gaia, producing an excess of ``twin'' wide binary population with $\Delta G<0.25$ \citep{El-Badry2019,Hwang2022twin}. After we remove nearby stars by the criterion of parallax$<0.5$\,mas, the binary contributions are strongly suppressed and the completeness correction is well-behaved.

\begin{figure*}
\centering
  \includegraphics[width=0.98\textwidth]{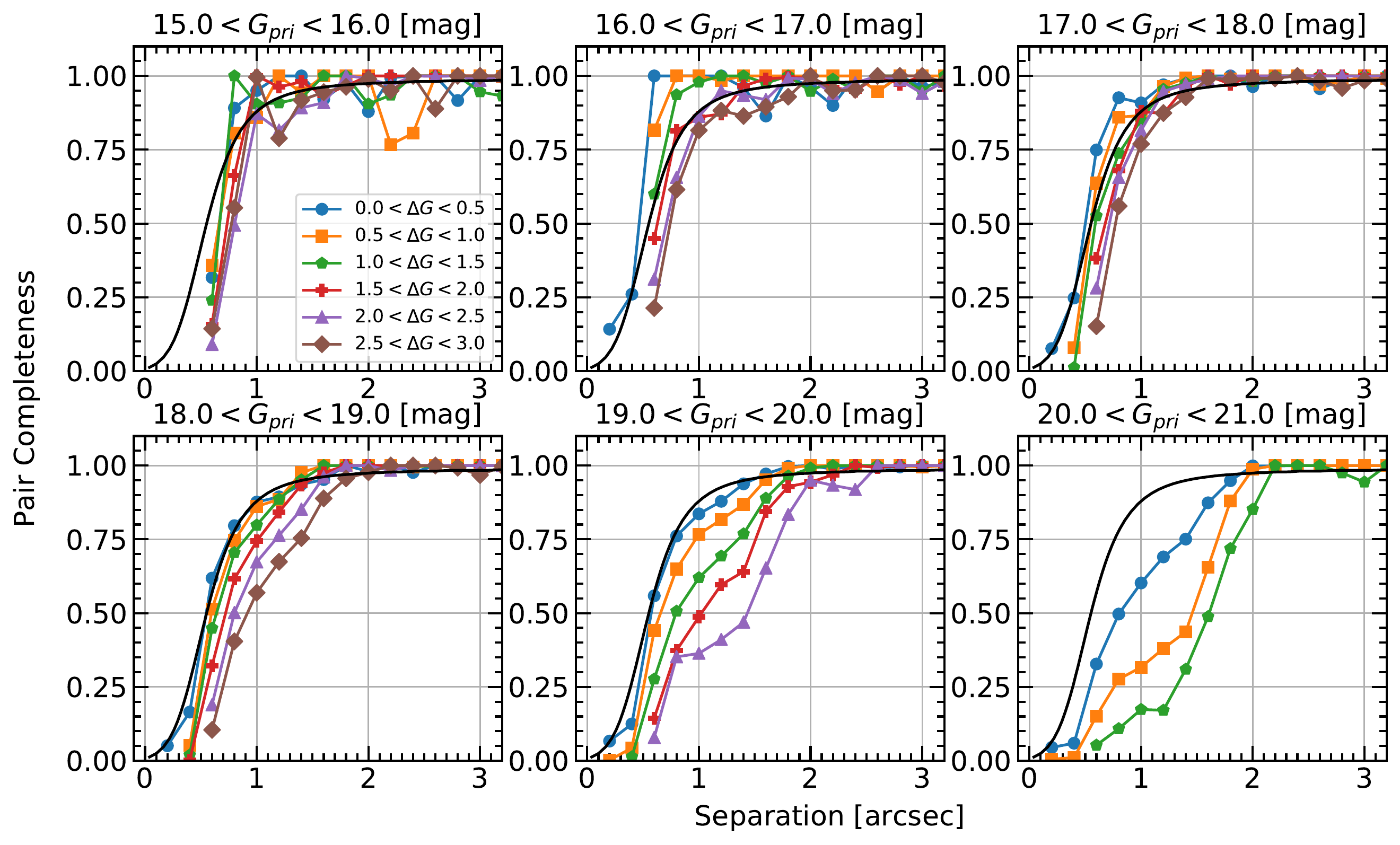}
  \caption{Pair completeness as functions of separation, primary magnitude and magnitude contrast $\Delta G$ estimated using the approach in \S\ref{sec:pair_comp}. The black line is the completeness measured using the WDS catalog from \citet{Fabricius_etal_2021}. 
    \label{fig:pair_comp}}
  \end{figure*}

Fig.~\ref{fig:pair_comp} shows the pair completeness as a function of angular separation. Each panel represents a different range of the primary's G-band magnitude $G_{pri}$, and each colored symbol is for different magnitude difference $\Delta G$. For a handful of bins the completeness correction can slightly exceed unity due to Poisson fluctuations, and we manually set $f_{\rm comp}$ to 1 in these bins (our quasar pairs rarely fall in these bins anyway). The black lines in Fig.~\ref{fig:pair_comp} are the model-fitted completeness of Gaia EDR3 derived from \citet{Fabricius_etal_2021} using the Washington Double Star Catalog \citep{Mason_etal_2001}, which does not take primary magnitudes and magnitude differences into account. The top panels show that our completeness correction for bright $G_{pri}$ bins agrees well with the black line. At the fainter end of $G_{pri}>18$, however, our completeness correction shows that $\Delta G$ plays an important role which is not captured by the black line, emphasizing the importance of deriving the customized completeness correction for our quasar sample. Due to Gaia's detection limit at $\sim21$\,mag, only $\Delta G<1.5$ have completeness correction available in the $20<G_{pri}<21$ panel (bottom right). The pair completeness correction on the 3D grid of ($G_{pri}$, $\Delta G$ and $\Delta\theta$) is available as an electronic FITS table with its content described in Table \ref{tab:comp}.


\begin{table}
\caption{Binned Pair Completeness}\label{tab:comp}
\resizebox{\columnwidth}{!}{%
\begin{tabular}{llll}
\hline\hline
Column & Format & Units & Description \\
(1) & (2) & (3) & (4) \\
\hline
GPRI  &     FLOAT[2]  & mag &   Boundary of $G_{pri}$   \\
DG & FLOAT[7] & mag & Boundaries of the $\Delta G$ grid  \\
DTHETA & FLOAT[20] & arcsec & Boundaries of the $\Delta\theta$ grid \\
FCOMP & FLOAT[19,6] & & Pair completeness  \\
\hline
\hline\\
\end{tabular}
}
{\raggedright Notes. For each row of the FITS table, GPRI is the boundary of $G_{pri}$, and DG and DTHETA are the boundaries (not bin center) of the $\Delta G$ and $\Delta\theta$ grids in that $G_{pri}$ bin, respectively. The bin size for the $\Delta G$ grid is 0.5 mag, and the bin size for the $\Delta\theta$ grid is 0\farcs2. The binned completeness FCOMP is set to ``NAN'' if the observed number of star pairs is $< 3$ in that bin. }
\end{table}

\begin{table}
\caption{Binned Quasar Pair Statistics}\label{tab:pair_stat}
\resizebox{\columnwidth}{!}{%
\begin{tabular}{lcccccc}
\hline\hline
$\Delta\theta$ (\arcsec) & $N_{\rm QQ}$ & $N_{\rm QQ, corr}$ & $\sigma_{-}$ & $\sigma_{+}$ & $\sigma_{\rm poisson}$ & $N_{\rm QQ, EDR3}$  \\
(1) & (2) & (3) & (4) & (5) & (6) & (7) \\
\hline
0.4 & 1 & 8.0 & 8.0 & 8.0 & 8.0 & 3.7 \\
0.6 & 9 & 21.0 & 6.7 & 6.7 & 7.0 & 15.8 \\
0.8 & 7 & 9.2 & 2.8 & 2.9 & 3.5 & 9.0 \\
1.0 & 4 & 4.7 & 2.4 & 2.4 & 2.4 & 4.6 \\
1.2 & 6 & 6.9 & 3.4 & 2.3 & 2.8 & 6.5 \\
1.4 & 6 & 6.5 & 2.2 & 3.1 & 2.7 & 6.3 \\
1.6 & 1 & 1.0 & 1.0 & 1.0 & 1.0 & 1.0 \\
1.8 & 5 & 5.0 & 2.0 & 2.0 & 2.2 & 5.2 \\
2.0 & 3 & 3.0 & 2.0 & 2.0 & 1.7 & 3.1 \\
2.2 & 6 & 6.1 & 2.1 & 2.1 & 2.5 & 6.1 \\
2.4 & 4 & 4.0 & 2.0 & 2.0 & 2.0 & 4.1 \\
2.6 & 2 & 2.0 & 1.0 & 1.0 & 1.4 & 2.0 \\
2.8 & 3 & 3.0 & 2.0 & 1.0 & 1.7 & 3.1 \\
3.0 & 3 & 3.0 & 2.0 & 2.0 & 1.7 & 3.1 \\
0.3--3.1 & 60 & 83.5 & 7.1 & 7.5 & -- &  73.5 \\
\hline
\hline\\
\end{tabular}
}
{\raggedright Notes. Pair statistics are measured in $\Delta\theta$ bins with a linear bin size of $0\farcs2$. Columns (3)--(5) are the pair statistics corrected for completeness ($N_{\rm QQ, corr}$), with the uncertainties ($\sigma_{-}$ and $\sigma{+}$) estimated from bootstrap resampling. Column (6) lists the uncertainties in $N_{\rm QQ, corr}$ estimated from Poisson counting uncertainties from the raw pair counts $N_{\rm QQ}$. Column (7) lists the corrected pair counts using the estimated completeness in \citet{Fabricius_etal_2021}, which remains less than unity even at $\Delta\theta>2\arcsec$. }
\end{table}

\section{Results}\label{sec:result}

\begin{figure}
\centering
  \includegraphics[width=0.48\textwidth]{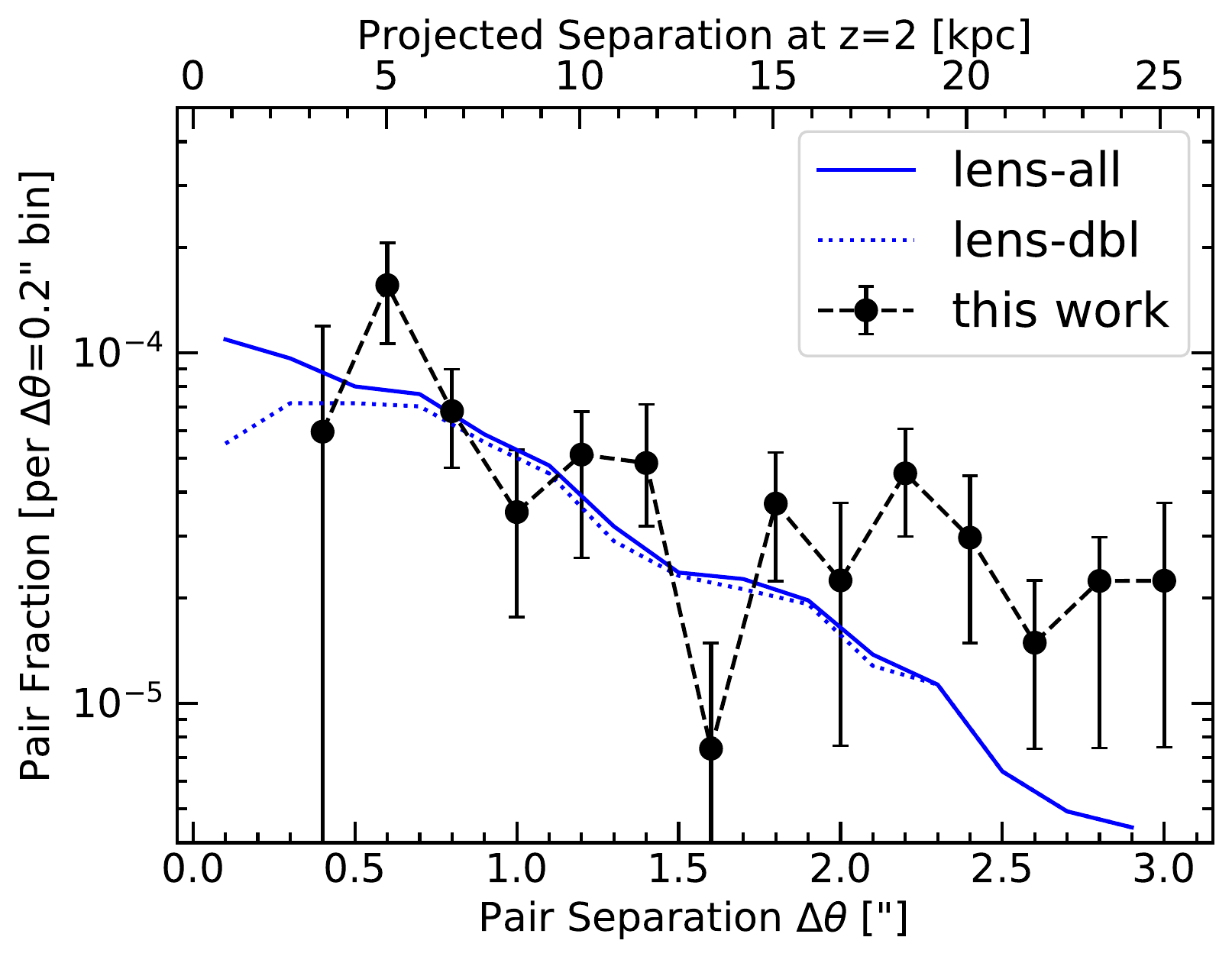}
  \caption{ Measured quasar pair fraction at $z>1.5$ and $G<20.25$ (black circles with error bars), corrected for pair incompleteness (\S\ref{sec:pair_comp}). Pair fraction uncertainties are estimated using bootstrap resampling. For comparison, we show the theoretically predicted lensed quasar fraction (matched in redshift and flux limit to our observed sample) from an updated version of the mock catalog described in \citet{Oguri_Marshall_2010}, for all lenses (solid blue line) and doubly-lensed quasars (dotted blue line), respectively.
    \label{fig:pair_stat}}
  \end{figure}

\begin{figure}
\centering
  \includegraphics[width=0.48\textwidth]{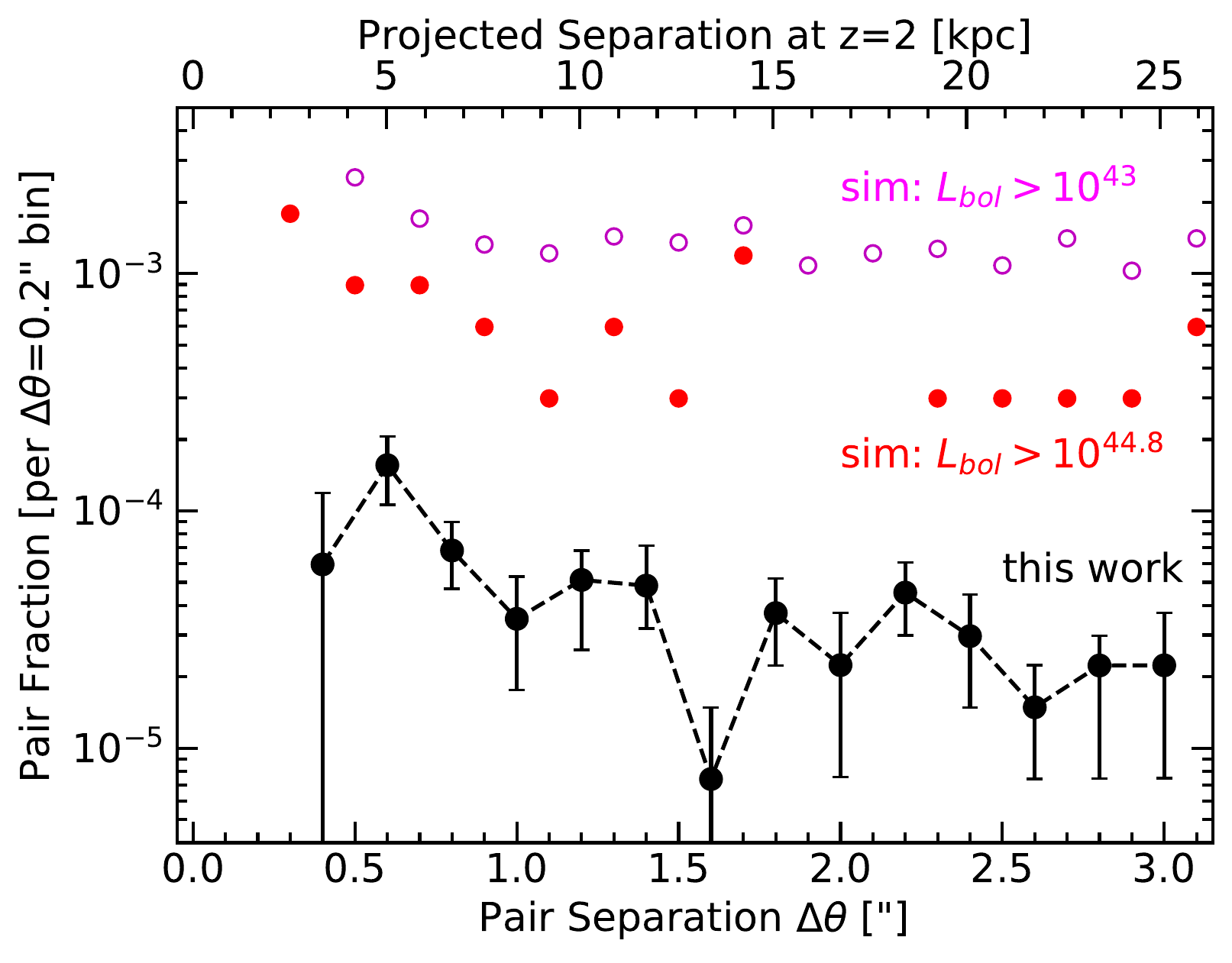}
  \caption{Same as Fig.~\ref{fig:pair_stat} but for the comparison between observed AGN pair statistics (black circles with error bars) and predictions from the \ASTRID\ cosmological simulation (cyan and red points) at $z\sim 2$ \citep[][]{ChenN_etal_2022c}. The simulated sample only has statistics to probe AGNs that are at least ten times fainter than our quasar sample, and includes both unobscured and obscured AGNs. 
    \label{fig:pair_stat_sim}}
  \end{figure}

To calculate the pair fraction, we define the parent quasar sample as the $\sim 134$\,k SDSS quasars with the same magnitude and redshift cuts as our pair sample, but are unresolved by Gaia. The overall abundance of double quasars with separations over $\sim 0\farcs3-3\arcsec$ is negligible compared to the parent single quasar population (e.g., even after completeness correction, the total pair fraction over these scales is of order $10^{-4}$ to $10^{-3}$). In any case, the parent sample only provides the denominator in the pair fraction calculation and does not affect the relative fraction as a function of pair separation.    

We show the completeness-corrected double quasar fraction as a function of angular separation in Fig.~\ref{fig:pair_stat}, where the pair fraction is defined as the ratio between the number of pairs in each separation bin and the total number of quasars in the parent sample. In detail, we use the binned completeness estimates $f_{\rm comp}$ over a grid of G magnitude of the primary, magnitude contrast and angular separation as quantified in \S\ref{sec:pair_comp}. Each quasar pair in our sample is weighted up by $1/f_{\rm comp}$, and the correction is significant only in the sub-arcsec regime. We estimate the uncertainties of the corrected pair statistics using bootstrap resampling of the pairs, which are consistent with the Poisson uncertainties estimated from the raw pair counts in each separation bin (see Table~\ref{tab:pair_stat}). The cumulative pair fraction within $0\farcs3-3\farcs1$ is $6.2\pm0.5\times 10^{-4}$ among $z>1.5$ quasars. Dividing these quasar pairs at their median redshift $\langle z \rangle =2$, we measure overall pair fractions over these scales of $6.6\pm1.2\times 10^{-4}$ and $5.9\pm1.0\times 10^{-4}$ in the lower ($\langle z \rangle =1.7$) and higher ($\langle z \rangle =2.4$) redshift bins, respectively, indicating there is no strong evolution in the pair fraction over the redshift range probed by our sample. 


For completeness, we also present in Table~\ref{tab:pair_stat} the corrected pair statistics using the pair-resolving completeness estimated in \citet{Fabricius_etal_2021} based on the WDS catalog. As demonstrated in \S\ref{sec:pair_comp}, our quasars are substantially fainter than sources in the WDS catalog, and the pair completeness in the sub-arcsec regime is somewhat lower than that in \citet{Fabricius_etal_2021}. Nevertheless, the corrected pair statistics using the completeness in \citet{Fabricius_etal_2021} are consistent with our fiducial estimates within $1\sigma$. The uncertainty in the pair statistics is the largest in the smallest $\Delta\theta$ bin, where there is only one observed quasar pair at 0\farcs4 separation, J0841+4825. This particular pair was first reported in \citet{Shen_etal_2021NatAs} as a genuine double quasar, although their data were insufficient to rule out the lensed quasar scenario. 

As shown in Fig.~\ref{fig:pair_stat}, the completeness-corrected double quasar fraction (per linear separation bin) gradually rises towards smaller separations (for reference, 1\arcsec\ corresponds to $\sim 8.5$\,kpc at $z\sim 2$). A constant pair fraction with separation is consistent with a quasar auto-correlation function of $\xi(r)\propto (r/r_0)^{-2}$, where $r$ is the 3-dimensional pair distance and $r_0$ is the correlation length. The steepening of the pair fraction towards small separations implies a steepening in small-scale quasar clustering at $\lesssim 30\,$kpc physical scales. 


Not all pairs in our final cleaned sample of 60 systems are physical quasar pairs, as some of them should be gravitationally lensed quasars. The lensed quasar statistics in the sub-arcsec regime is not well constrained observationally, and therefore we use an updated mock catalog of lensed quasars from \citet{Oguri_Marshall_2010} to estimate the lensed quasar contribution \citep[see also][]{Lemon_etal_2022}. The updated version uses a galaxy velocity dispersion function for all types of galaxies (in contrast to only early-type galaxies considered in the original version of the mock) and imposes no lower limit on the image separation (in contrast to the lower limit of image separation of $0\farcs5$ in the original version of the mock).

We include all lensed systems (e.g., doubles, quads, etc.) in the mock catalog with two (and only two) images above the flux limit, with the same redshift cut of $z>1.5$ as for our observed sample. The mock catalog uses SDSS $i$ band magnitude, and we adopt $i<20.2$ for individual resolved images that roughly corresponds to the same $G$-band limit used for the observed pair sample. Varying the flux limit in the mock lensed quasar catalog by one magnitude introduces less than a factor of two in the lensed quasar fractions. Lensed quasars with more than two images above the flux limit would not have been included in our pair sample. The lensed quasar fraction (blue lines in Fig.~\ref{fig:pair_stat}) also shows a gradual increase towards the sub-arcsec regime, mainly due to the increase in the abundance of less massive lens galaxies. 


In Fig.~\ref{fig:pair_stat_sim}, we compare our double quasar fraction with predictions for dual AGNs at $z\sim 2$ in the cosmological hydrodynamic simulation \texttt{ASTRID} \citep{ChenN_etal_2022c}. \texttt{ASTRID} is a recently developed large-volume, high-resolution (with a gravitational softening of $1.5\,{\rm kpc}/h$ and a dark matter mass
resolution of $9.6\times 10^6\,M_\odot$) cosmological hydrodynamic simulation that studies the evolution of galaxies and SMBHs. It utilizes a new version of the \texttt{MP-Gadget} \citep{MPGadget2018} simulation code to solve the gravitational evolution (with an N-body tree-particle-mesh approach), hydrodynamics (with Smoothed Particle Hydrodynamics), and astrophysical processes with a series of subgrid models. With a comoving volume of $(250\,{\rm Mpc}/h)^3$, \texttt{ASTRID} is the largest galaxy formation simulation up to date that covers the epoch of the cosmic noon. The large volume of \texttt{ASTRID} can provide a statistical sample of the rare quasar population, and the high resolution enables detailed studies of the quasar pair statistics and environments down to galactic scales. Details of the \texttt{ASTRID} simulation and the SMBH population overview can be found in \citet{Bird2022,Ni2022,ChenN2022b}, and a comprehensive analysis of the dual AGN population predicted by \texttt{ASTRID} can be found in \cite{ChenN2022b,ChenN_etal_2022c}.

Given the simulation volume of \ASTRID, we can only explore dual AGNs with lower luminosities than our quasar sample. For instance, there are 3 dual AGNs at $z\sim 2$ in \ASTRID\ that sample the same luminosity and pair separation ranges as our observed sample, which is not enough for detailed statistical analysis. We therefore use two lower bolometric luminosity cuts, $L_{\rm bol}>10^{44.8}\,{\rm erg\,s^{-1}}$ and $L_{\rm bol}>10^{43}\,{\rm erg\,s^{-1}}$, to select the parent single AGNs and dual AGNs in \ASTRID. We impose a BH mass cut of $M_{\rm BH}>10^7\,M_\odot$ in the simulated AGNs -- this BH mass scale is resolved in \ASTRID. The simulated dual AGNs are restricted to have radial separations $<50\,{\rm kpc}$ and transverse separations $<30\,{\rm kpc}$. The resulting simulated dual AGN sample includes 59 and 1282 pairs for the two luminosity cuts, respectively. As shown in Fig.~\ref{fig:pair_stat_sim}, the dual AGN fraction is generally lower for the higher luminosity cut. Nevertheless, both simulated samples show an enhancement in the pair fraction towards the smallest separations, as seen in the observed sample.
 
If we could increase the luminosity threshold further in the simulated AGN sample as pair statistics allow, we would expect to see further reduced dual fraction. This luminosity trend in the dual AGN fraction can be qualitatively understood as follows: if assuming no merger-enhanced AGN duty cycles $f(L_{\rm min})$, the dual AGN fraction (among all AGNs with the same luminosity threshold $L_{\rm min}$) depends on $L_{\rm min}$ through the duty cycle, i.e., $\propto f(L_{\rm min})$, while the fraction of pairs with a single AGN among all AGNs is constant with the luminosity threshold. As $L_{\rm min}$ increases, the duty cycle $f(L_{\rm min})$ decreases, leading to reduced dual AGN fraction among all AGNs. The average AGN duty cycle at $z\sim 2$ in the \ASTRID\ simulation (for all $M_{\rm BH}>10^7\,M_\odot$ SMBHs) roughly decreases by a factor of 10 from $L_{\rm bol}>10^{43}\,{\rm erg\,s^{-1}}$ to $L_{\rm bol}>10^{44.8}\,{\rm erg\,s^{-1}}$, and by another factor of $\sim 10$ from $L_{\rm bol}>10^{44.8}\,{\rm erg\,s^{-1}}$ to $L_{\rm bol}>10^{45.8}\,{\rm erg\,s^{-1}}$ (N.~Chen et~al., in prep). Thus we expect the simulated dual AGN fraction for $L_{\rm bol}>10^{45.8}\,{\rm erg\,s^{-1}}$ (matching our observed sample) would be a factor $\sim 10$ smaller than the red solid points in Fig.~\ref{fig:pair_stat_sim}, which would match the observed statistics. However, we do expect somewhat enhanced AGN duty cycles in galaxy mergers, which would elevate the simulated dual AGN fraction.

The above comparison with simulations should be interpreted with some caution. First of all, currently the simulated sample does not distinguish between obscured and unobscured AGNs, while our quasar sample only contains unobscured objects. Simulations at $z\sim2$ have shown that many luminous dual AGNs are completely obscured in gas-rich mergers \citep{ChenN_etal_2022c}, which would significantly reduce the observable fraction of unobscured dual quasars. Secondly, it might be necessary to further match the SMBH masses in this comparison, i.e., the observed SDSS quasars have BH masses $>$ a few $\times 10^8\,M_\odot$ \citep[e.g.,][]{Shen_etal_2011}. In any case, we conclude that the observed double quasar statistics are roughly consistent with predictions for the dual AGN population from simulations.

The intriguing finding that the observed double quasar statistics are consistent with theoretical predictions for both lensing and simulated dual AGNs may indicate that these two populations are comparable in number by coincidence. A complete division between lenses and dual quasars in our sample with follow-up observations will fully address this important issue. We further discuss the implications of our observed pair statistics in \S\ref{sec:disc}. 

Our definition of the quasar pair fraction is free of a selection bias related to the flux limit and source blending. When selecting a sample of unresolved systems (either single quasars or unresolved pairs), potential pairs or lensed images would boost the combined flux to above the flux limit, and enhance the presence of pairs in the parent sample. In case of gravitational lenses, this is the magnification bias. However, our Gaia sample is a resolved pair sample, and each component of the pair is above the flux limit. In other words, our pair fraction is defined as the fraction of $G<20.25$ quasars that have a resolved quasar companion that is also brighter than $G=20.25$. Pairs with either of the components fainter than the flux limit, even if the other component or the combined flux is above the flux limit, would not have been included in our pair sample to contribute to the numerator. The parent quasar sample has the same flux limit, and could include fainter pairs or lensed images that boost the combined fluxes above the threshold, but such small-scale pairs/lensed images are rare and would only slightly perturb the denominator in our pair fraction calculation. 

\begin{figure}
\centering
  \includegraphics[width=0.48\textwidth]{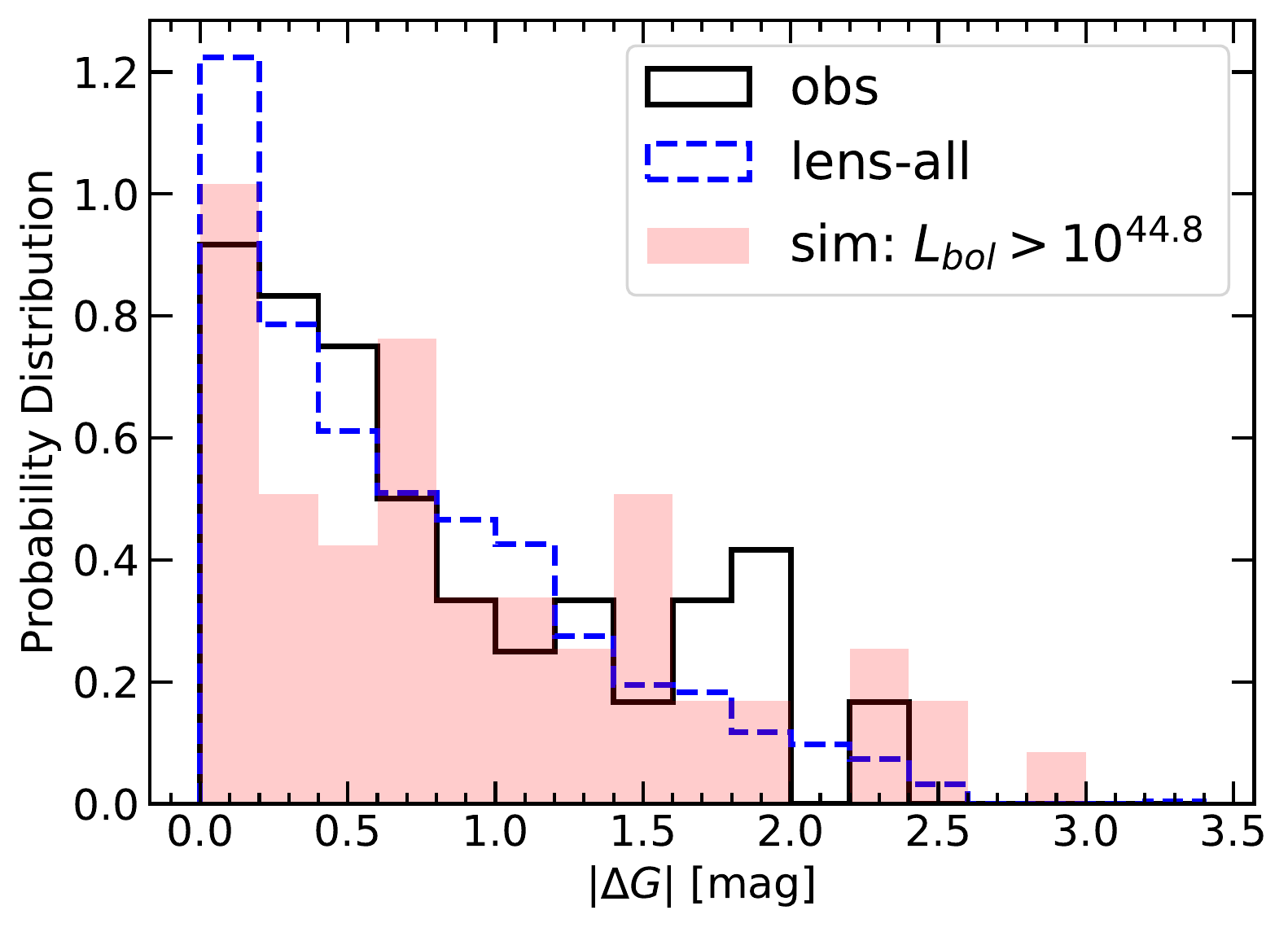}
  \caption{Pair flux ratio (or magnitude contrast) distribution of the 60 double quasars in our sample (black solid histogram). The predictions for lensed quasars using the mock catalog in \citet{Oguri_Marshall_2010} and for dual AGNs in the cosmological hydrodynamic simulation \ASTRID\ \citep[][]{ChenN_etal_2022c} are shown in the blue dashed and pink shaded histograms, respectively. 
    \label{fig:flux_ratio}}
\end{figure}

We next examine the flux ratios of the observed quasar pairs. The sample statistics is insufficient to explore the flux ratios as a function of separation in detail, and hence we focus the discussion on the distribution for the full quasar pair sample with $\Delta\theta<3\arcsec$. However, dividing the quasar pair sample into wide-separation ($\Delta\theta>1\arcsec$) and close-separation ($\Delta\theta<1\arcsec$) pairs, there is no noticeable difference in the pair flux ratio distribution. Fig.~\ref{fig:flux_ratio} shows that the observed quasar pair flux ratio distribution peaks near unity. 

\begin{figure}
\centering
  \includegraphics[width=0.48\textwidth]{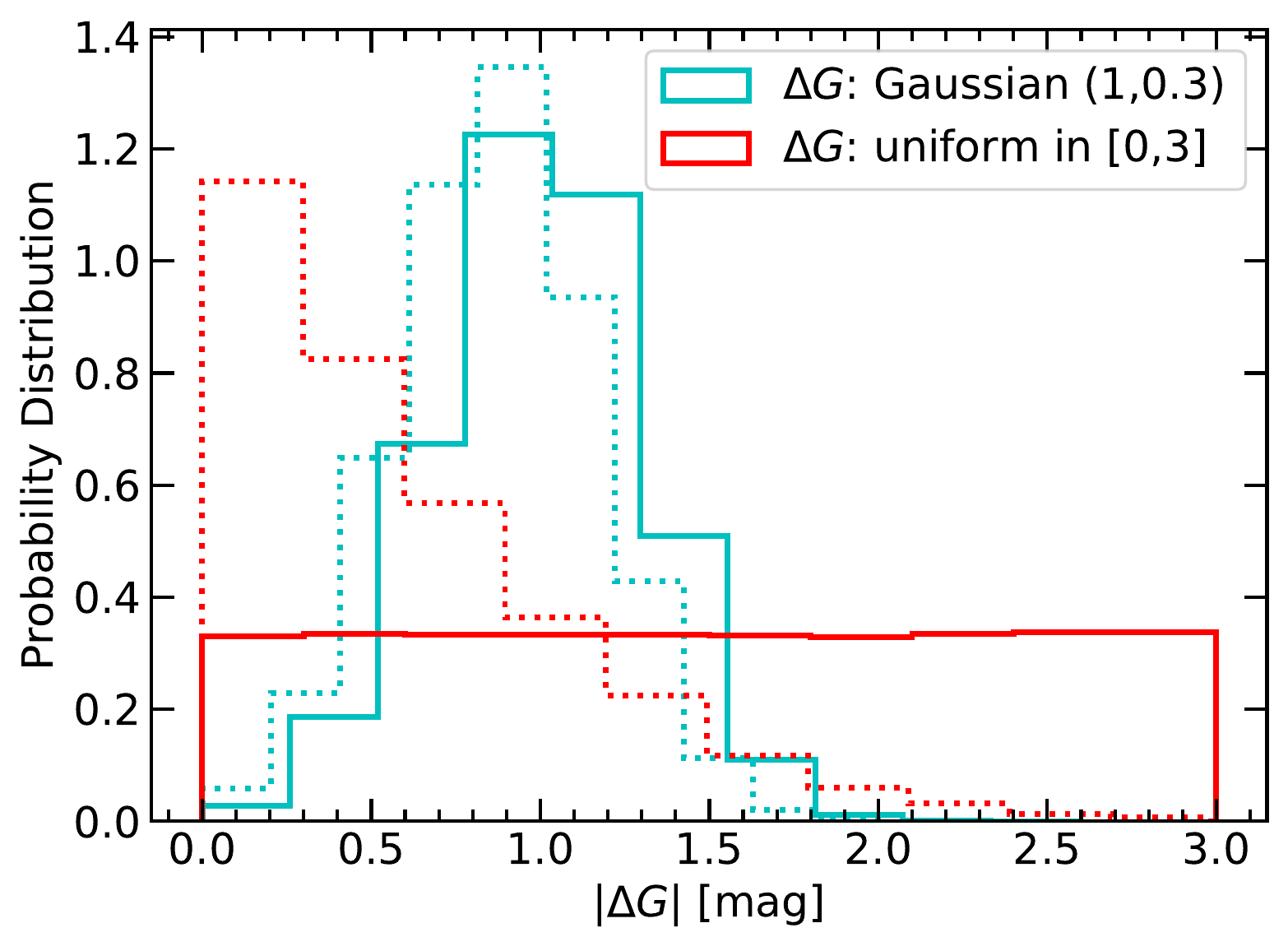}
  \caption{Effects of the flux limit ($G<20.25$) on the observed pair flux ratio distribution. The magnitude distribution of the primary quasar follows the observed distribution. The cyan and red solid histograms show two intrinsic flux ratio distributions: (1) a Gaussian distribution of $\Delta G$ with mean and dispersion of 1 and 0.3 mag; (2) a uniform distribution of $\Delta G$ over [0,3] mag. The dotted histograms show the resulting observed flux ratio (magnitude contrast) distribution in each case. The flux limit imposed on both components of the pair enhances the prominence of the peak towards equal flux ratio. If the intrinsic flux ratio distribution is broad (i.e., in the uniform distribution case), selection effects due to the common flux limit would produce a strong peak near equal flux ratio. 
    \label{fig:flux_ratio_selection}}
\end{figure}

The flux limit ($G<20.25$) in our sample selection reduces the dynamic range in the flux ratio of the observed pairs, biasing the distribution of flux ratios towards more equal-flux values. To illustrate this effect, we consider the wide-separation ($1\arcsec<\Delta\theta\lesssim 3\arcsec$) quasar pairs in our sample, since this subset does not suffer from pair-resolving completeness as much as the close-separation pairs do (\S\ref{sec:pair_comp}). In other words, the main selection effect is due to the flux limit $G<20.25$ for both components of the pair. In Fig.~\ref{fig:flux_ratio_selection} we demonstrate the effect of the flux limit with simple, idealized simulations, with an assumed intrinsic flux ratio ($\Delta G$) distribution (solid lines). The magnitude distribution of the primary (brighter) component follows the observed distribution (Fig.~\ref{fig:pair_prop}). We test two different input $\Delta G$ distribution: (1) a Gaussian distribution with mean $\Delta G=1$ (mag) and a dispersion of 0.3 mag, and (2) a uniform distribution of $\Delta G$ within [0,3]\,mag. When the input $\Delta G$ distribution narrowly peaks at a non-equal ratio value  (the Gaussian distribution case), the observed pair flux ratio distribution (dotted lines) is slightly shifted to smaller values, but the peak is more or less preserved. On the other hand, if the input distribution is broad (the uniform distribution case), the resulting observed $\Delta G$ distribution peaks at equal flux ratio.




In Fig.~\ref{fig:flux_ratio}, we show the AGN pair flux ratio distribution from the \ASTRID\ simulation at $z\sim 2$ \citep[][]{ChenN2022b}, again restricting to physical AGN pairs with transverse separations $r_p<30\,{\rm kpc}$ and radial separations $<50\,{\rm kpc}$. Because the AGN population in the simulations is limited by the simulation volume, we relax the luminosity threshold for both components to be $L_{\rm bol}> 10^{44.8}\,{\rm erg\,s^{-1}}$. The pair flux ratios from the simulated AGN pairs also peak around equal-flux ratio, although the peak is somewhat less prominent than that of the observed sample. If we lower the luminosity threshold in the simulated sample, more pairs with large luminosity ratios will be included in the sample, further weakening the peak prominence at equal-flux ratio. Similarly, if we increase the luminosity threshold to match our sample ($L_{\rm bol}>10^{45.8}\, {\rm erg\,s^{-1}}$), we expect simulated dual AGNs would produce a prominent peak around unity flux ratio, similar to the observed distribution. The intrinsic pair flux ratio distribution (for the SMBH pair population) from the simulations, however, is much broader if we relax the flux limit on the fainter component. Synchronized growth of the pair of SMBHs that rapidly drives their masses towards equality does not seem to be the case on these $<$ tens of kpc scales.

Finally, we show the flux ratio of lensed quasar images in Fig.~\ref{fig:flux_ratio}, using the same mock catalog described above. Coincidently, flux ratios of lensed quasar images also peak around unity flux ratio. This peak is primarily due to selection effects. Double lenses with large magnification factors (which tend to have small magnitude contrasts) from intrinsically fainter quasars are over-represented in the flux-limited sample due to the lensing magnification bias. Quad lenses often have two bright images with similar magnifications near the critical curve and this population preferentially resides in the small-separation regime. The magnification bias would also enhance the presence of equal-flux quad lenses (only the two brightest images) in the flux-limited sample. Therefore, the observed pair flux ratio distribution cannot be used to readily distinguish the lensing and quasar pair scenarios in the statistical sense.





\section{Discussion}\label{sec:disc}

Since we focus on luminous quasars at $z>1.5$ with bolometric luminosities $L_{\rm bol}>10^{45.8}\,{\rm erg\,s^{-1}}$ and the bulk of the sample are near the flux limit, we make the assumption that the intrinsic quasar pair fraction (as a function of separation) is more or less constant over the luminosity range probed in this work. This simplifies most of the following discussions. The luminosity dependence of pair statistics will be explored in future work with improved sample statistics and more extended dynamic range in quasar luminosity. 

\subsection{Lensing vs Pairs}\label{sec:disc1}

Fig.~\ref{fig:pair_stat} shows that the measured quasar pair fraction in our Gaia sample agrees well with the predicted lensed quasar population at $<1\farcs5$ separations. However, we caution that the lensed quasar fraction is based on mock catalogs and may be different from the actual lensed quasar population at these small separations. In reality, the observed quasar pairs are comparable in number to lensed quasars over few \arcsec\ separations \citep[e.g.,][]{Hennawi_etal_2006,Hennawi_etal_2010,Kayo_Oguri_2012}, but the relative numbers between lensed quasars and pairs are unconstrained at $\lesssim 1\arcsec$. There is also an observational bias that preferentially removes sub-arcsec lensed quasars from the oberved pair sample: lensed quasars are associated with lensing galaxies, which could (if the lensing galaxy is at $z\lesssim 1.5$) change the optical colors of the unresolved system and reduce the probability of selection from ground-based surveys such as the SDSS. Therefore we expect at least some of these double quasars are genuine pairs rather than lenses. This is indeed the case, as there are already several confirmed quasar pairs in our sample from the literature, as discussed in \S\ref{sec:sample}.

Follow-up observations of the full sample of 60 double quasar systems will conclusively reveal the division between lensed quasars and physical pairs over the full range of $\sim 0\farcs3-3\arcsec$ separations. It is notoriously difficult to distinguish these two scenarios at high redshift and small separations \citep[e.g.,][]{Shen_etal_2021NatAs,Yue_etal_2021}. Minor spectral dissimilarities between the two components of the pair are insufficient to rule out lensing \citep[e.g.,][]{Shen_etal_2021NatAs}, while spectral similarities are equally insufficient to rule out a quasar pair since different quasars can look similar in their spectral appearances \citep[e.g.,][]{Rochais_etal_2017}, particularly at $z>1.5$ where optical spectroscopy only covers the rest-frame UV broad lines. Spatially-resolved near-IR spectroscopy may be able to reveal the differences in the narrow emission lines, e.g., \OIII, in a quasar pair, but is challenging given the S/N requirement and relatively weak narrow-line emission in high-$z$, high-luminosity quasars \citep[e.g.,][]{Shen_2016}. 

Multi-wavelength coverage of the two resolved components may help reject the lensing scenario, if the spectral energy distributions are markedly different, e.g., with additional high-resolution radio imaging of the resolved pair. The most decisive and efficient observation to rule out lensing, however, is probably the non-detection of a potential lens galaxy in deep imaging. In the case of $z>1.5$ candidate quasar pairs at sub-arcsec separations, this test requires high spatial resolution and deep IR imaging, ideally from HST or JWST. Indeed, existing optical imaging data (even taken with HST) are too shallow to rule out the lensing hypothesis \citep[e.g.,][]{Shen_etal_2021NatAs,Chen_etal_2022} for $z>1.5$ double quasars, even though statistics may slightly favor the dual quasar scenario over lensing \citep{Shen_etal_2021NatAs}. High-$z$ lens galaxies will be faint in the optical, and the non-detection limit of the lens placed by HST optical imaging is not stringent enough. For larger-separation pairs, deep IR imaging from ground would be sufficient to rule out (or confirm) the lensing scenario based on the non-detection (or detection) of a lens galaxy. Deep IR imaging may also be able to reveal tidal features in the host of the pair, offering additional evidence for physical merging pairs. 


In what follows, we remain agnostic about the division between lensing and pairs in our sample, and discuss different outcomes if one or the other population dominates our pair sample.

\subsection{Dynamical friction, quasar duty cycles, and recoiling SMBHs}

The overall double quasar fraction from our sample, $f_{QQ}\sim 6\times 10^{-4}$ ($r_p\sim 3-30$\,kpc) among all $G<20.25$ quasars at $\langle z\rangle\approx 2$, is lower than the dual AGN fraction ($\sim 10^{-2}$) at similar redshifts and separations predicted from recent hydrodynamic simulations \citep[e.g.,][and references therein]{Hirschmann_etal_2014,Dubois_etal_2014,Steinborn2016,Rosas-Guevara_etal_2019,DeRosa_etal_2019,Volonteri_etal_2022}. The main reason for this apparent discrepancy is due to the fact that these simulations do not have sufficient volume to probe the most luminous quasars, and focus on the much less luminous AGN population ($L_{\rm bol}>10^{43}\,{\rm erg\,s^{-1}}$). These low-luminosity AGNs have much higher duty cycles than luminous quasars. In general, the dual AGN fraction among AGNs increases as the luminosity threshold decreases, as seen in the simulations (Fig.~\ref{fig:pair_stat_sim}), as well as the observed high dual AGN fraction ($\gtrsim$ few percent) among low-luminosity AGNs in the nearby Universe \citep{Liu_etal_2011,Koss_etal_2012}. Improvements in both the observed sample (to fainter flux limits) and in the simulation volume over the next few years will enable a better comparison. 

A substantial fraction of AGNs in these simulations are also optically obscured, and would not be included in our sample. If obscuration occurs more often in merging pairs than in single AGNs, the dual AGN fraction for the unobscured population will be reduced compared with that for all AGNs. Even if the obscured fraction is the same among single AGNs and AGNs in pairs, requiring both AGNs in the pair to be unobscured would also lead to a reduced dual AGN fraction for unobscured AGNs (similar to the duty cycle argument).

On the observational side and focusing on quasar luminosities ($L_{\rm bol}\gtrsim 10^{45}\,{\rm erg\,s^{-1}}$), \citet{Kayo_Oguri_2012} reported a dual quasar fraction of $\sim 5\times 10^{-4}$ over $0.6<z<2.2$ and $10\lesssim r_{p}\lesssim 100\,{\rm kpc}$, which is roughly in line with our measured double quasar fraction over smaller separations and higher redshifts. On the other hand, using ground-based optical imaging of resolved pairs around SDSS quasars from the Hyper Suprime-Cam Subaru Strategic Program, \citet{Silverman_etal_2020} reported a double quasar fraction (dual and lensed quasars combined) of $0.26\pm0.18\%$ (requiring a pair flux ratio $>0.1$) over $r_p=3-30\,{\rm kpc}$ with no redshift evolution, which is a factor of $\sim 4$ higher (albeit still within $\sim 1\sigma$) than our pair fraction over the same separations. There is a slight difference in the selection of double quasars between our work and \citet{Silverman_etal_2020}: while the flux limit of the primary SDSS quasar is the same, we require the companion is also brighter than this flux limit, while \citet{Silverman_etal_2020} includes companions that can be ten times fainter than the primary SDSS quasar. Therefore we expect some of the double quasars (candidates) in \citet{Silverman_etal_2020} would not pass our selection. 

Furthermore, the \citet{Silverman_etal_2020} measurement is based on a ground-based imaging pair sample with loose color selection of quasars, and spectroscopic follow-up is required to remove foreground star contamination in these apparent pairs, as acknowledged by \citet{Silverman_etal_2020}. Our earlier results based on HST imaging and spectroscopic follow-up of high-redshift candidate quasar pairs have shown that such stellar contamination is significant (e.g., $>50\%$) for pure photometric color selection \citep{Chen_etal_2022}. Foreground star contamination would also be a problem in other predominantly imaging samples of dual/offset AGN candidates \citep{Stemo_etal_2021}. In our SDSS+Gaia approach, the additional proper motion information and the rejection of foreground star superpositions with spectral PCA delivered a much cleaner double quasar sample.

The relative frequency of quasar pairs as a function of separation in the $r_{p}\sim 3-30\,{\rm kpc}$ regime is determined by the dynamical friction timescale and the duty cycle of quasar activity in mergers, both of which are functions of separation. If the quasar duty cycle remains constant over these separations, simple prediction from dynamical friction implies a roughly constant pair fraction per linear separation bin towards smaller separations \citep[e.g.,][]{Yu_2002,Chen_etal_2020b}. 

If the pair statistics shown in Fig.~\ref{fig:pair_stat} are dominated by physical quasar pairs, then the rising pair fraction (per linear separation bin) towards small separations indicates that the quasar duty cycle is elevated towards smaller separations, or that the dynamical friction timescale deviates from the scaling predicted in analytical calculations by, e.g., \citet{Chen_etal_2020b}. The observed rising quasar pair fraction towards small separations for our high-redshift sample is consistent with observations at low redshift, where the AGN pair fraction also increases towards small separations at $r_{p}\lesssim 30\,{\rm kpc}$ \citep[e.g.,][]{Ellison_etal_2011,Liu_etal_2012,Stemo_etal_2021}. Such an elevation of SMBH accretion at small pair separations, i.e., late stages of galaxy mergers, is also seen in some hydrodynamic simulations \citep[e.g.,][]{Capelo_etal_2017}.

On the other hand, if the pair statistics shown in Fig.~\ref{fig:pair_stat} are dominated by lensed quasars, and the intrinsic physical pair fraction is flat or even decreasing towards smaller separations, it would imply little enhanced (or even reduced) quasar activity towards the $\sim {\rm kpc}$ regime in galaxy mergers, which would be at odds with numerical simulation results and low-redshift observational results. Alternatively, it may imply that at $z\sim 2$, pairs of SMBHs decay more rapidly towards the $\sim {\rm kpc}$ regime than predicted by dynamical friction from stars, for example, accelerated by the presence of gas \citep[e.g.,][]{Callegari_etal_2009} expected in high-redshift gas-rich mergers, or by the build-up of a dense nuclear stellar cusp around one or both SMBHs \citep[e.g.,][]{VanWassenhove_etal_2014}. 

Either way, our sample of 60 double quasars can be used to address these different scenarios and constrain the dynamical friction evolution of the SMBH pair, as well as the duty cycle of quasar activity in mergers. To that end, we are conducting follow-up observations to differentiate the pairs versus lensing scenarios for our sample, and will present the results in future work.  

We end this section by pointing out the possibility that a tiny fraction of these quasar pairs might contain an accreting recoiled SMBH from the prior merger of two SMBHs \citep[e.g.,][and references therein]{Blecha_etal_2016}. However, there are still significant theoretical uncertainties on this putative population of recoiling SMBHs and observational challenges to distinguish them from insprialing SMBHs in galaxy mergers. Perhaps host galaxy properties can be useful to identify recoiling SMBHs as offset AGNs, e.g., if these rogue SMBHs predominately reside in early type galaxies long after the merger. 


\subsection{Fuzzy Dark Matter and a Possible $\sim$\,kpc Pile-up}

In the fuzzy dark matter (FDM) model and ignoring baryonic effects, SMBH pairs in galaxy mergers will stall at $\lambda_{\rm FDM}\sim$\,kpc scales due to energy injection from fluctuations of dark matter particles on their de Broglie wavelength $\lambda_{\rm FDM}$ \citep[e.g.,][]{Hui2017}. If the duty cycle of quasar activity is independent of pair evolution, we expect to see a dramatic pile up of quasar pairs near the stall distance, because these pairs spend much longer time there (i.e., $\sim$ Hubble time) compared to their lifetime during previous galactic inspiral. Our quasar pair sample does not yet well probe the $<1\,$kpc regime, and we do not observe any sudden spike in the pair fraction towards $\sim 0\farcs2$ (corresponding to $\sim 1.6\,{\rm kpc}$ at $z\sim 2$). Pair statistics with future data sets (see \S\ref{sec:con}) will probe the sub-kpc regime and constrain the nature of FDM. 

However, absence of evidence is not evidence of absence. The potential lack of a pile up of quasar pairs below $\sim 1\,$kpc can be explained by baryonic effects, i.e., the pair orbit can further decay regardless of the energy pumping from FDM fluctuations. In addition, in the final stage of pair evolution long after the initial galaxy merger, accretion onto SMBHs may become much less efficient, leading to a diminished fraction of dual quasars among these stalled $\sim$\,kpc SMBH pairs. The most exciting aspect of this test is to potentially reveal that there is indeed a pile up of quasar pairs on $\lesssim$\,kpc scales, which would offer strong support to the FDM model. Compared to other observational tests \citep{Hui2017}, the statistics of $\sim$kpc-scale quasar pairs offer a simple but potentially definitive test (but see below), hinging on the discovery of such a pile-up of SMBH pairs.  

On the other hand, certain dynamical processes associated with baryonic matter might also lead to the stalling of SMBH pairs at $\sim$kpc scales \citep[][and references therein]{lisa22}. For example, in the case of massive (e.g., $>10^7\,M_\odot$) SMBHs, clumpiness in the host galaxy and inhomogeneous gas and stellar density profiles can lead to inefficient inspiral and potentially stalling of the SMBH pair at $\sim$kpc separations \citep[e.g.,][]{Tamburello_etal_2017,Pfister_etal_2019,Bortolas_etal_2020}. This is still in early theoretical investigations, and observations of host galaxies of high-redshift dual quasars might offer insights on these dynamical processes.


\section{Conclusions}\label{sec:con}

In this work we have measured the quasar pair statistics over $\sim 0\farcs3-3\arcsec$ separations at $z>1.5$ (median redshift $\langle z \rangle\approx 2$), using a sample of 60 resolved double quasars from Gaia EDR3 \citep{Fabricius_etal_2021}. These pairs are selected by cross-matching the Gaia EDR3 catalog with spectroscopically confirmed quasars from SDSS DR16 \citep{Lyke_etal_2020}. Both members of the pair are flux limited to $G<20.25$, therefore our pair sample corresponds to the luminous quasar population at cosmic noon, with $L_{\rm bol}>10^{45.8}\,{\rm erg\,s^{-1}}$ at $z>1.5$. We efficiently separate quasars and stars in resolved pairs using Gaia proper motion measurements and PCA analysis of SDSS spectra (\S\ref{sec:sample}). We quantify the pair completeness in Gaia EDR3 as functions of pair separation $\Delta\theta$, magnitude of the primary, and magnitude contrast of the pair (\S\ref{sec:pair_comp}). The completeness-corrected pair fraction (per linear separation bin; among all $z>1.5$ quasars at $G<20.25$) increases towards smaller separations, and is elevated by a factor of $\sim 5$ from $\Delta\theta\sim 3\arcsec$ to $\Delta\theta\sim 0\farcs3$. The integrated pair fraction over $\sim 0\farcs3-3\arcsec$ scales (corresponding to projected physical separations of $\sim 3-30\,{\rm kpc}$ at $z\sim 2$) is $\sim 6.2\pm0.5\times 10^{-4}$, with no obvious evolution in the redshift range of our sample.

The major caveat of the current analysis is that the division between physical quasar pairs and gravitationally lensed quasars is unknown, especially in the sub-arcsec regime. Previous searches of high-redshift quasar pairs and lensed quasars on $>1\arcsec$ scales have revealed that both populations contribute significantly to the observed double quasars \citep[e.g.,][]{Hennawi_etal_2006,Hennawi_etal_2010,Myers_etal_2008,Kayo_Oguri_2012,More_etal_2016,Eftekharzadeh_etal_2017}. It is then reasonable to expect that there are both bona fide quasar pairs and lensed quasars in the sub-arcsec regime. We are conducting follow-up observations for the complete sample of 60 double quasars presented here, and will refine our constraints on the quasar pair statistics.  

This work represents a meaningful advance on observational constraints on the formation and evolution of SMBH pairs at high redshift. Granted, the depth of Gaia and SDSS limits such a systematic search to the most luminous quasars, missing the bulk of rapidly growing SMBHs at cosmic noon. The important and more abundant populations of single offset AGNs in mergers and obscured AGNs are also not explored with the Gaia+SDSS sample. Nevertheless, this approach with Gaia+SDSS has delivered some of the first statistical measurements of quasar pair fraction in a redshift-separation regime that has just started to be explored in a systematic fashion \citep[e.g.,][]{Silverman_etal_2020,Stemo_etal_2021,Chen_etal_2022}. With continued Gaia observations (more resolved pairs from different scanning directions), we expect to recover additional luminous quasar pairs at $z>1.5$ to improve the statistics. 

However, the intrinsic abundance of luminous quasar pairs at cosmic noon is low. In order to significantly improve the pair statistics, to extend to lower AGN luminosities, and to explore the diversity in SMBH and host properties, it is necessary to carry out similar searches with deeper, wide-area surveys at sub-arcsec resolution. Upcoming wide-field space missions, such as Euclid \citep[][to be launched in $\sim 2023$]{Euclid_EWS}, the Chinese Space Station Telescope \citep[CSST,][to be launched in $\sim 2024$]{Zhan_2021}, and the Nancy Grace Roman Space Telescope \citep[Roman,][to be launched before $\sim 2027$]{Spergel_etal_2015}, will provide the perfect combination to perform systematic searches of SMBH pairs across cosmic time. All three missions will carry out a wide-field imaging survey in multiple filters with $\sim 0\farcs05-0\farcs2$ resolution and depths of $\sim 25-28$ AB mag, with additional spectroscopic capabilities. The combined photometric data cover a broad wavelength range across UV-optical-near-infrared. These data can be used to efficiently select candidate quasar pairs based on photometric colors and spectroscopic information, down to the diffraction limit of these space telescopes. Dedicated follow-up observations of these candidates can confirm the nature of these pairs, if needed. 

In particular, the deep IR imaging from Euclid and Roman will be useful to test the lensing scenario for high-redshift double quasars. In addition, the capability of detecting the host galaxy in deep IR imaging and measuring sub-arcsec offset of point sources within will enable the systematic discovery of single offset AGNs in high-redshift mergers. Host galaxy measurements will also allow a detailed look at the populations of dual and offset AGNs in different types of galaxies, shedding light on AGN fueling and recoiling SMBHs. With combined data sets from these upcoming space-based surveys, we will conclusively measure the abundances of galactic-scale quasar and AGN pairs, offset AGNs, and sub-arcsec lensed quasars across most of the cosmic history, with unprecedented statistics and coverage of the parameter space of SMBHs and host galaxies.

\acknowledgments

We thank P. Capelo and Joe Hennawi for useful discussions. This work is partially supported by NSF grants AST-2009947 (YS) and AST-2108162 (YS, XL). MO acknowledges support by JSPS KAKENHI Grant Numbers JP22H01260, JP20H05856, JP20H00181. HCH acknowledges the support of the Infosys Membership at the Institute for Advanced Study. NLZ acknowledges support by NASA through grant HST-GO-15900 from the Space Telescope Science Institute and by the Institute for Advanced Study through J. Robbert Oppenheimer Visiting Professorship and the Bershadsky Fund. This research was supported in part by the National Science Foundation under Grant No. PHY-1748958. We are grateful to the hospitality of the Kavli Institute for Theoretical Physics at UC Santa Barbara and the KITP Conference: Building Bridges: Towards a Unified Picture of Stellar and Black Hole Binary Accretion and Evolution (May 2022) during which part of the work was performed.  

\bibliography{refs}

\end{document}